# Overpotential decomposition enabled decoupling of complex kinetic processes in battery electrodes

Ruoyu Xiong[1], Yue Yu[1], Shuyi Chen[1], Maoyuan Li[1], Longhui Li[1], Mengyuan Zhou[1], Wen Zhang[2], Bo yan[3], Dequn Li[1], Hui Yang[2,*], Yun Zhang[1,*], Huamin Zhou[1,*]

[1]State Key Laboratory of Material Processing and Die & Mould Technology, School of Materials Science and Engineering, Huazhong University of Science and Technology, Wuhan 430074, China

[2]Department of Mechanics, School of Aerospace Engineering, Huazhong University of Science and Technology, Wuhan, Hubei 430074, China

[3]School of Materials Science and Engineering, Shanghai Jiao Tong University, Shanghai, 200030, China

## Abstract

Identifying overpotential components of electrochemical systems enables quantitative analysis of polarization contributions of kinetic processes under practical operating conditions. However, the inherently coupled kinetic processes lead to an enormous challenge in measuring individual overpotentials, particularly in composite electrodes of lithium-ion batteries. Herein, the full decomposition of electrode overpotential is realized by the collaboration of single-layer structured particle electrode (SLPE) constructions and time-resolved potential measurements, explicitly revealing the evolution of kinetic processes. Perfect prediction of the discharging profiles is achieved via potential measurements on SLPEs, even in extreme polarization conditions. By decoupling overpotentials in different electrode/cell structures and material systems, the dominant limiting processes of battery rate performance are uncovered, based on which the optimization of electrochemical kinetics can be conducted. Our study not only shades light on decoupling complex kinetics in electrochemical systems, but also provides vitally significant guidance for the rational design of high-performance batteries.

Keywords: overpotentials; kinetic processes; battery electrodes; rate performance; single-layer particle electrode

## Introduction

Overpotential is a fundamental physical quantity in the research of various electrochemical systems, such as batteries, fuel cells, and electrolyzers. It stems from the polarization of charge/species transfer in the bulk or at the interface of electrode/electrolyte [1,2] and stands as the fingerprint of these kinetic processes. Acquiring overpotential components is an effective manner to capture complex kinetic processes, providing direct bases for analyzing, designing, and optimizing the system in energy efficiency [3,4], power performance [5], heat generation [6], and aging [7,8]. However, measuring individual overpotentials of each process is always challenging, particularly for porous composite electrodes with ion-intercalation materials in lithium-ion batteries (LIBs). There are four processes in the composite electrode domain during battery operating [9,10]: i) electron conduction on the composite matrix of the electrode; ii) $Li^+$ transport in the liquid electrolyte filled in the pores; iii) charge transfer reaction at the particle-electrolyte interface; iv) solid diffusion of Li (polaron) within the active particles. These kinetic processes are inherently coupled, resulting in an entanglement of various overpotentials in the measured voltage/potential signal [11,12]. The overpotentials also vary in the electrode thickness direction and evolve over time because of the concentration and potential gradients induced by directional species and charge transport through porous electrodes [13-15]. As a result of the complicated characteristics of entanglement and spatial-temporal distribution, individual overpotentials have never been directly measured. Previous researches have been conducted from modeling perspectives [5,16-18], while the obtained overpotential components are non-validated, leading to extensive debate about which processes limit battery voltage and capacity in fast charging/discharging [19-21]. Meanwhile, abundant experiments have revealed that, the rate performance of LIBs can be improved to some extent by modulating the material and structure of electrodes, including active particles [22,23], matrix pores [24-26], conductive network [27-29], electrolyte [30,31], and surface/interface [32-34], however, a systematic understanding of various reaction/transport processes has not yet been reached.

In this work, the full decomposition of electrode overpotential is realized by the collaboration of single-layer structured particle electrode (SLPE) constructions and time-resolved potential measurements. The innovation of SLPE effectively eliminates the limitations of electron conduction and liquid ion transport from all kinetic processes tangled in conventional electrodes (CEs), leading to the isolation of the interfacial charge transfer and solid diffusion. The isolated two are further decoupled via time-resolved potential measurements (1 ms) in three-electrode configurations. With the determination of the Ohmic voltage drop caused by electron conduction, the total overpotential involved in CEs can be finally decomposed. The proposed method is verified in typical LIBs material systems, including $LiNi_{0.5}Mn_{0.3}Co_{0.2}O_2$ (NMC532), $LiCoO_2$, and graphite. The evolutions of overpotential components under different operating conditions are uncovered, from which the rate-determining kinetic processes are identified for further electrode/cell design and optimization.

## Decoupling strategy

As shown in Figure 1, the total overpotential of the CE is composed of individuals caused by electron conduction in the composite matrix ($\eta_e$), $Li^+$ transport in the liquid electrolyte ($\eta_l$), charge transfer reaction at the particle-electrolyte interface ($\eta_{ct}$), and solid diffusion of $Li^+$ (polaron) within the active particles ($\eta_s$). To determine each component of the overpotential in CE, an SLPE structure is designed in the first step. As there are only single-layer active particles sparsely distributed on a metallic current collector with sufficient conductive agents, it is reasonable to assume that $\eta_e$ can be ignored owing to excellent conduction. In addition, the electrolyte overpotential $\eta_l$ also can be neglected because of the excess electrolyte surrounding each active particle and the short ion transport path from the separator. Therefore, according to the SLPE structure design above, $\eta_e$ and $\eta_l$ can be eliminated, leaving $\eta_{ct}$ and $\eta_s$ for further decomposition.

To decouple the overpotentials $\eta_{ct}$ and $\eta_s$, time-resolved potential measurements with millisecond resolution are conducted on SLPEs under a step current stimulation. Because of apparent differences in response time of interfacial charge transfer and solid diffusion within active particles, the broadly confirmed frequency-domain characteristics of the two processes (1-10 Hz vs. < 0.1 Hz) by electrochemical impedance spectrum (EIS) [35-38] are also supposed to be reflected in the time domain as different responses in potential. Considering frequency-domain methods can neither reveal the non-linear performances of $\eta_{ct}$ and $\eta_s$ at nonequilibrium states (i.e., at practical charge/discharge rates) nor capture their dynamic evolution as the lithiation/delithiation proceeds [39-41], the time-domain potential measurements with a high sampling frequency (1 kHz) are adopted to distinguish the dynamic behaviors of $\eta_{ct}$ and $\eta_s$ in the designed SLPE structure. It should be noted that, if the potential measurements are directly performed on CEs, there would be a non-negligible overpotential contribution from liquid ion transport ($\eta_l$), of which the characteristic time scale overlaps those of charge transfer and solid diffusion [35,41].

Assuming that the overpotential contributions from the interfacial charge transfer ($\eta_{ct}$) and solid diffusion within active particles ($\eta_s$) obtained from the SLPE can be mapped into the CE with the consideration of the inhomogeneity of electrochemical reaction in the thickness direction, for a full decomposition of the total overpotential of the CE, the influences of electron conduction in the composite matrix ($\eta_e$) and $Li^+$ transport in the liquid electrolyte ($\eta_l$) need to be further identified. Since the $\eta_e$ can be regarded as the Ohmic voltage drop, it can be directly obtained from the resistance in EIS plots of the CE. With the measurement of the total overpotential of the CE, the last remaining component $\eta_l$ can be calculated by subtracting the other three parts from the total overpotential.

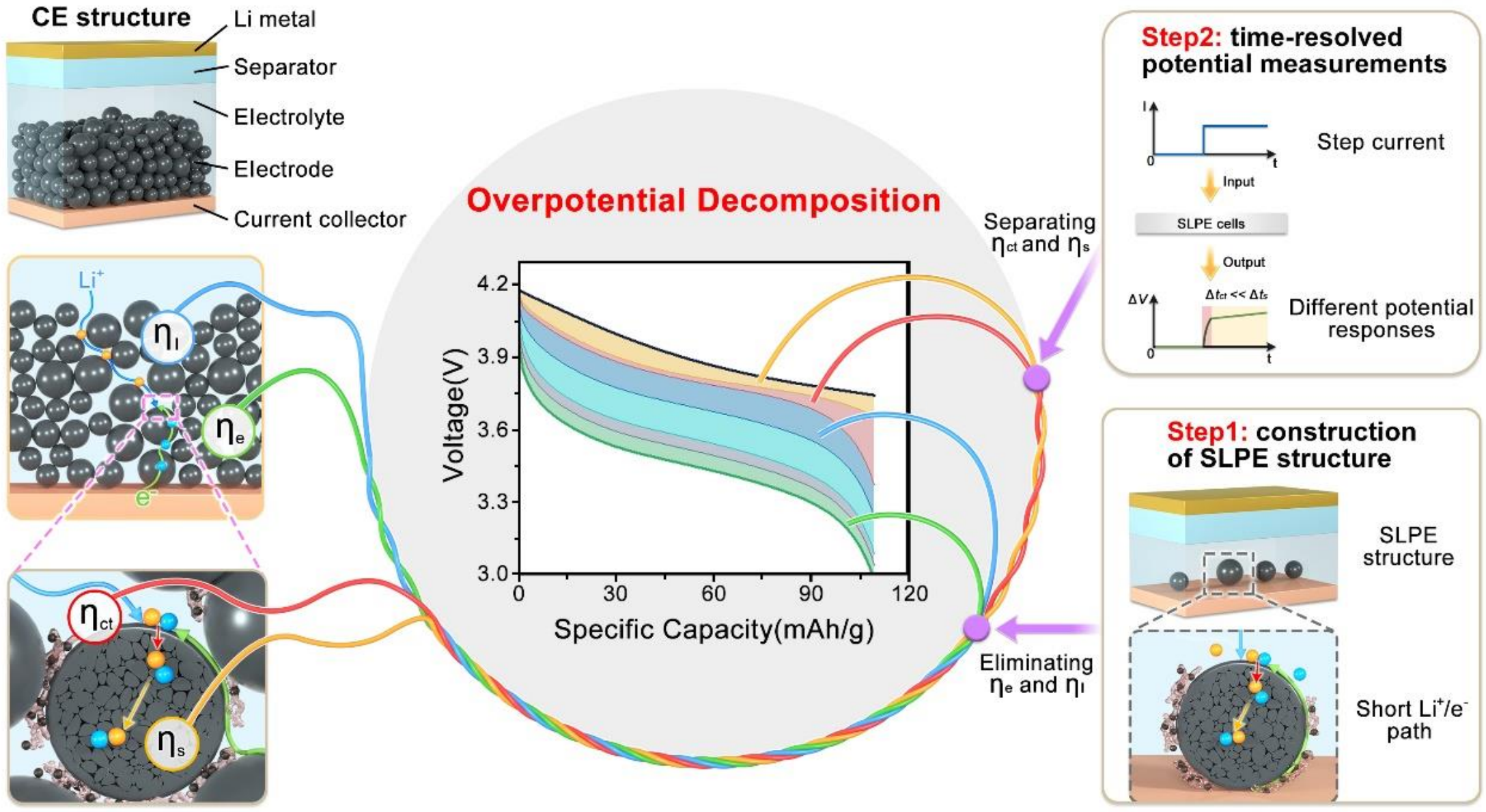

**Figure 1 Schematical illustration of overpotential decomposition strategy.** For the total overpotential of CEs caused by electron conduction in the composite matrix ($\eta_e$), $Li^+$ transport in the liquid electrolyte ($\eta_l$), charge transfer reaction at the particle-electrolyte interface ($\eta_{ct}$), and solid diffusion of $Li^+$ within the active particles ($\eta_s$), the first step is to eliminate $\eta_e$ and $\eta_l$ via the construction of SLPE structure, leaving $\eta_{ct}$ and $\eta_s$ for further decomposition. Under a step current stimulation, time-resolved potential measurements with millisecond resolution are conducted on SLPEs to distinguish the overpotentials of $\eta_{ct}$ and $\eta_s$ due to their different response time scales. Since the $\eta_e$ can be regarded as the Ohmic voltage drop, it can be directly obtained from the resistance in EIS plots of the CE. With the measurement of the total overpotential of the CE, the last remaining component $\eta_l$ can be calculated by subtracting the other three parts from the total overpotential.

## Construction of SLPEs

The SLPEs are fabricated via ultrathin slurry-casting with precise coating spacing control (Figure 2a and Methods). Figure 2b shows the SEM images of the SLPE structure, where single-layered $LiNi_{0.5}Mn_{0.3}Co_{0.2}O_2$ (NMC532) particles are sparsely distributed on the aluminum foil. SLPEs of $LiCoO_2$ and graphite are also demonstrated in Figure S1. For a specific SLPE, the packing density $n_A$, namely the number of active particles per unit area, can be defined as (the derivation details are provided in Equation S1-S3)

$$n_A = \frac{3m_{ap}l}{4\pi \overline{R_{ap}}^3 \rho_{ap} V} \tag{1}$$

where $m_{ap}$ and $\rho_{ap}$ denote the total mass and the specific mass of active particles in the suspension, respectively, $l$ is the coating gap, $R_{ap}$ the equivalent radii of active particles, and $V$ the volume of the suspension. By precisely controlling the coating gap $l$, the packing density can be effectively regulated. As shown in Figure 2c (corresponding SEM images shown in Figure S2), the packing densities measured from experimental results are closely consistent with that predicted by Equation 1. The as-prepared SLPEs together with Li metal anodes and commercial separators, soaked in an electrolyte solution, are then assembled into coin cells for further measurements. Basic electrochemical characterizations, including specific capacity, cyclic voltammetry, rate discharging, and cycling, demonstrate that the SLPE is a reliable electromechanical structure for kinetic measurements of active particles (Figure S3).

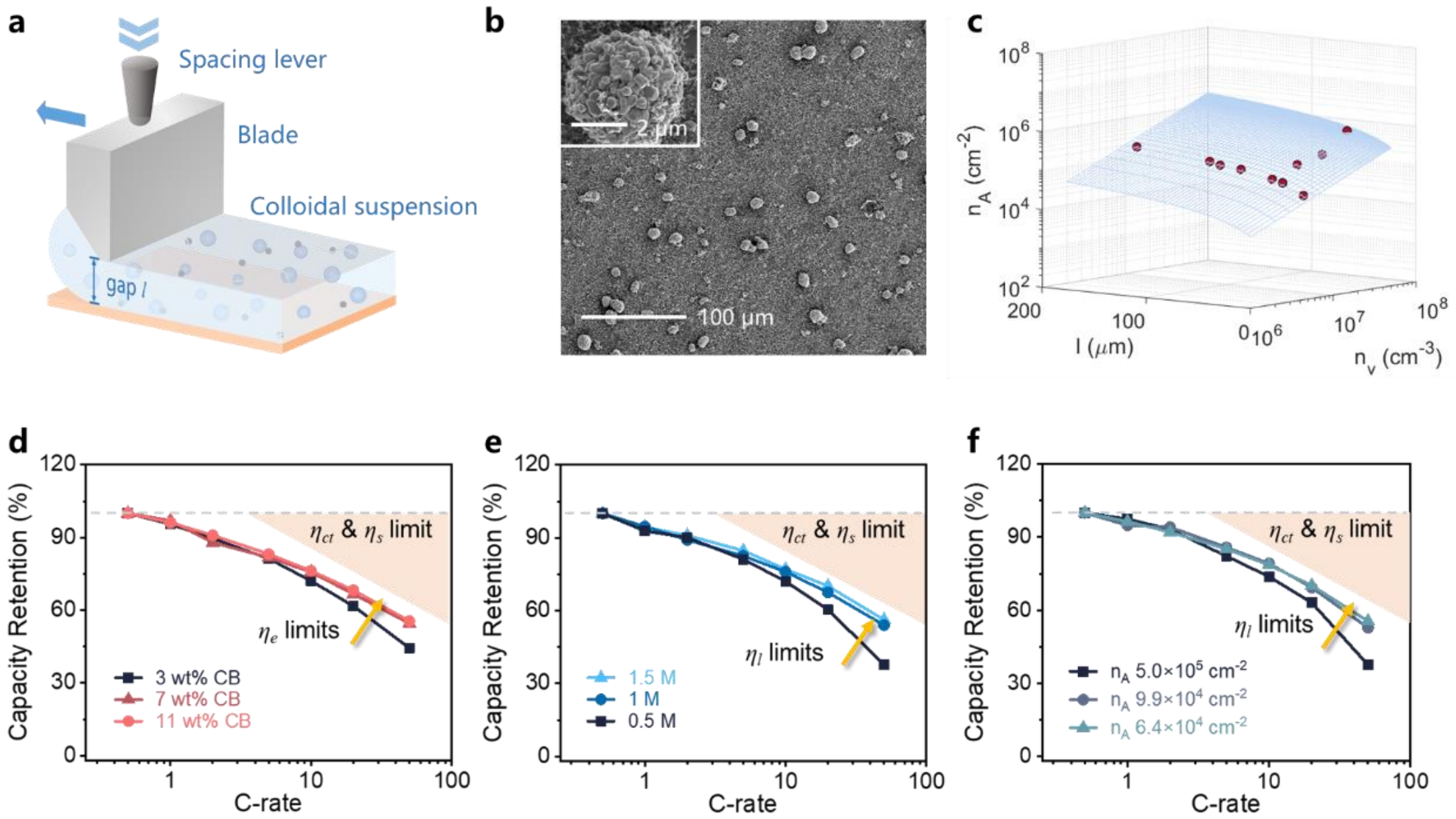

**Figure 2 Preparation and rate-performance testing of SLPEs.** a, Fabrication of SLPEs via ultrathin slurry-casting with precise coating spacing control. b, SEM images of NMC532 SLPEs with the inset showing the surface morphology of active particles. c, Both experimental (red dots) and theoretical (blue surface) results indicate that the packing density $n_A$ of SLPEs is quantitatively controlled by the coating gap $l$ and the spatial distribution density $n_V$. d-f, Rate capability of the SLPEs with different contents of CB (d), Li salt concentrations of the electrolyte (e), and packing densities (f).

Figure 2d shows the capacity retention of SLPE-based coin cells with different contents of carbon black (CB). It is clear that the SLPE with a higher content (7 wt%) of CB displays better capacity retention than that with a lower CB content (3 wt%) as C-rate increases. However, further raising the content of CB from 7 wt% to 11 wt% does not cause any obvious improvement in the cell performance, as the electron conduction in the composite matrix is facile enough that the corresponding overpotential $\eta_e$ becomes so small that it is no longer the factor limiting the rate capability of the SLPEs, which demonstrates that with sufficient conductivity additive in the SLPEs, the overpotential of the electron conduction $\eta_e$ would be neglected. For the opposite extreme condition that the content of CB decreases to 0 wt%, the SLPE exhibits a large overpotential and little capacity (Figure S4) due to the intrinsic poor conductivity of NMC532 material [42].

By adjusting the $LiPF_6$ concentration in electrolyte and packing density of active material while fixing the content of CB at 7 wt%, the $Li^+$ transport in the liquid electrolyte can also be promoted to eliminate the influence of ion-transport overpotential $\eta_l$. As shown in Figure 2e, the capacity retention drops rapidly at 50C in the case with 0.5M $LiPF_6$, probably because of the lack of $Li^+$ near the surface of active particles. With the increment of the $LiPF_6$ concentration, the capacity-rate curves converge gradually with the capacity retention improved. The inconspicuous discrepancy between the curves with 1M and 1.5M $LiPF_6$ indicates slight concentration polarization and thus a small enough $\eta_l$. Moreover, the concentration polarization caused by scrambling for $Li^+$ among adjacent particles can be also excluded through an ultralow packing density of active particles in the SLPEs. As plotted in Figure 2f, the rate performance is still limited by the concentration overpotential $\eta_l$ in the case with a packing density $n_A$=5.0×$10^5$ $cm^{-2}$. When the packing density decreases from 9.9×$10^4$ $cm^{-2}$ to 6.4×$10^4$ $cm^{-2}$, no perceptible change of capacity retention is revealed, indicating that once a low enough packing density is regulated by precise control of coating gap, the influence of ion-transport overpotential $\eta_l$ can be regarded as negligible. Our electrochemical modeling and simulations (the mathematic description is given by Equation S9-S25) further demonstrate that $\eta_l$ of the SLPE (1M $LiPF_6$, $n_A$=7×$10^4$ $cm^{-2}$) is ignorable (the range of about 0.01 V vs. the total voltage drop of 0.8 V) even at 50C (Figure S5a), owing to a moderate $Li^+$ concentration gradient in the electrolyte (Figure S5b).

The above findings reveal that, with sufficient CB, lithium salt, and low enough packing density, all capacity-rate curves converge in Figures 2d-f, suggesting the SLPEs are free of the limitations of $\eta_e$ and $\eta_l$, and influenced only by the $\eta_{ct}$ and $\eta_s$ (the orange regions). In contrast, the commonly used thin electrodes in kinetic investigations [43,44], which have a much higher thickness and packing density of active particles than the SLPEs, are still not rigorous for kinetic parameter estimation due to non-negligible $\eta_l$.

## Time-resolved potential measurements

The charge transfer overpotential $\eta_{ct}$ and solid diffusion overpotential $\eta_s$ are further decomposed by time-resolved potential measurements on SLPEs. Figure 3a shows the potential response of an SLPE cell to a step current input, collected with a sampling interval of 1 ms. Once the step current is applied ($t$=1.980 s), the potential drops rapidly from equilibrium potential $E_{eq}$ until $t$≈2.2 s, after which a slow decrement is followed. Such slow decline of potential is widely believed to originate from the sluggish solid ion diffusion within the active particles [35-37]. The locally enlarged view of the rapid potential decrease region shows two distinct potential drop periods, separated by the inflection point at $t$=1.982 s. The first one contains potential responses of the electron conduction between the current collector and active particles ($\eta_c$) and interface reaction of Li metal anode ($\eta_{Li}$), as their intrinsic response times are less than 0.1 ms [45-47] and 3 ms [38] (Figure S6), respectively. The second one corresponds to the potential responses of charge transfer ($\eta_{ct}$), of which response time is on the order of 100 ms. By linearly extrapolating the potential curve in the second period of the rapid decrease region and the slow potential decline part, two intersection points, i.e., P1 and P2, can be identified on the vertical line $t$=1.980 s. The potential drop from the equilibrium potential $E_{eq}$ to P1 consists of $\eta_c$ and $\eta_{Li}$, and the potential difference between P1 and P2 equals $\eta_{ct}$. By subtracting the potential of P2 from any instant potential (i.e. $t$=12 s), the solid diffusion overpotential $\eta_s$ can be obtained, which accumulates with time while $\eta_c$+$\eta_{Li}$ and $\eta_{ct}$ remain nearly constant. When the step current is interrupted ($t$=271.980 s), the SLPE cell exhibits an instantaneous potential jump in response to the current interruption, followed by a slow increase process toward $E_{eq}$ (Figure 3b). Different from the foregoing case, the overpotentials decrease with time. Overpotential components $\eta_{Li}$+$\eta_c$, $\eta_{ct}$, and $\eta_s$ at this moment can be calculated by the similar decomposition strategy.

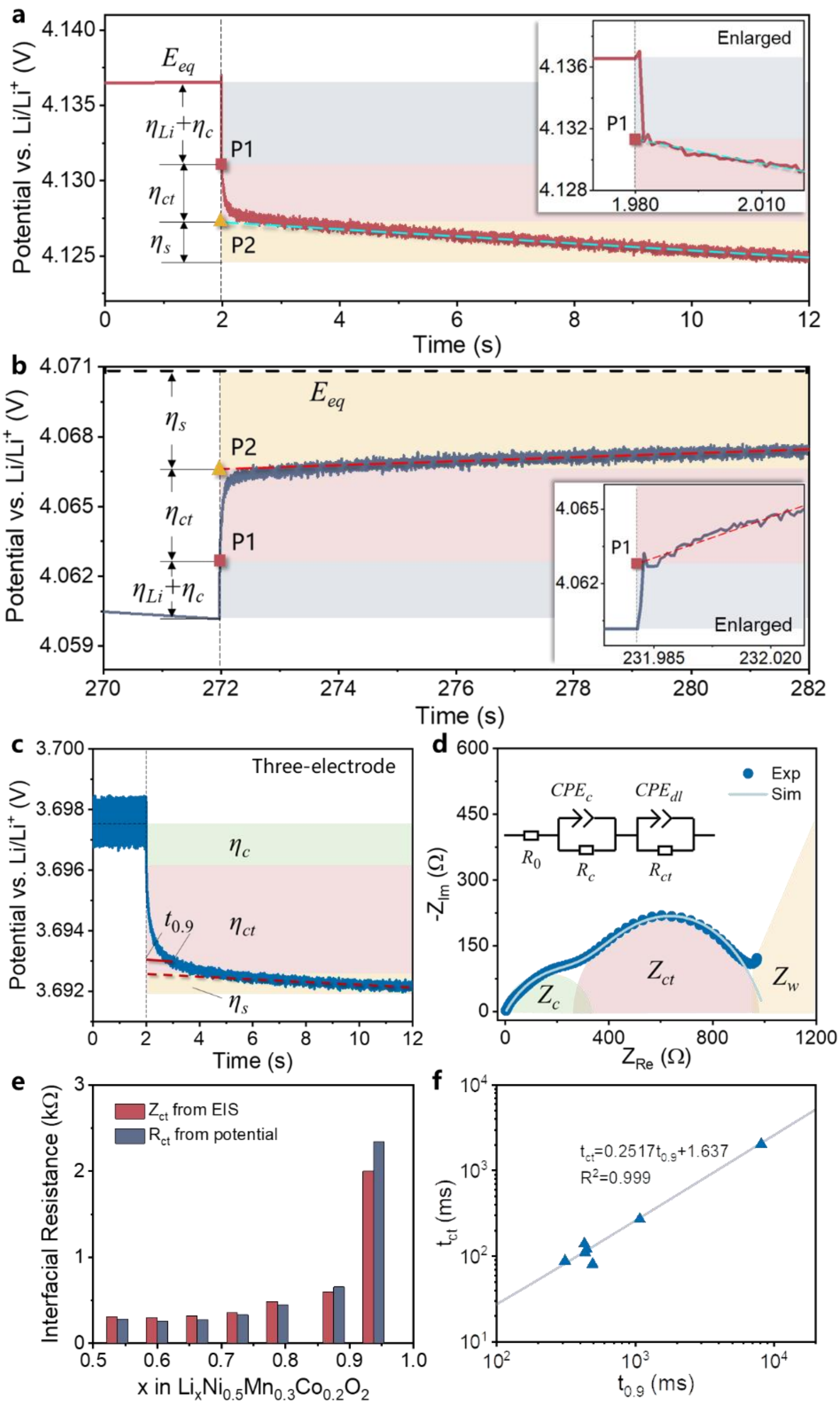

**Figure 3 Time-resolved potential measurements on SLPEs and validation.** a,b, Potential responses of an SLPE cell at the moment that a step current input is applied (a) and interrupted (b). When the current is loaded, the production of overpotentials $\eta_{Li}+\eta_c$, $\eta_{ct}$, and $\eta_s$ leads to the deviation of the potential from the equilibrium value $E_{eq}$; once the current is interrupted, the potential shits gradually toward $E_{eq}$ due to the reduction of these overpotentials. c, Potential response in the time domain at the beginning of the discharge stage from the piecewise discharge measurement in a three-electrode configuration. d, EIS

plots collected before discharging stage in (c). e,f, The comparison of the extracted charge transfer resistances (e) and characteristic times (f) from both the potential measurements and EIS, showing the feasibility of the time-resolved potential measurements.

The rationality of the time-resolved potential measurement is verified by EIS in a static (or quasi-static) condition. The potential sampling (1 ms) and impedance characterization (0.01 Hz - 1 MHz) of the SLPEs were conducted at various depths of discharge (DoD) by piecewise discharge in a three-electrode configuration (Figure S8), as shown in Figure S9. The EIS is collected after every discharging and resting stage of the piecewise discharge. Taking the case at 75% DoD as an example, the three stages of potential response in the time domain (Figure 3c) match the three impedance regions in the frequency domain (Figure 3d). In the Nyquist plots (Figure 3d), the smaller arc ($Z_c$) is the contact impedance of the current collector/electrode interface [45-47], and the bigger one ($Z_{ct}$), which increases with DoD (Figure S9a), corresponds to $\eta_{ct}$ because it depends on $Li^+$/vacancy concentration at the surface of active particles. The tail of the impedance curves (Warburg impedance $Z_w$) is associated with solid diffusion, which can be clearly distinguished at lower DoDs (Figure S9b). The equivalent circuit model illustrated in Figure 3d was adopted to fit the EIS data, in which constant phase elements (CPEs) instead of capacitors were used to describe the heterogeneous surface of active particles [47]. The resistances $Z_{ct}$ and time constants $t_{ct}$ of charge transfer extracted from fitting curves are then compared with the equivalent charge transfer resistance ($R_{ct}=\eta_{ct}/I$) and the response time $t_{0.9}$ (defined as the time when the potential reaches 90% of $\eta_c+\eta_{ct}$, as indicated in Figure 3c) from the time-resolved potential measurements. As shown in Figure 3e, $R_{ct}$ and $Z_{ct}$, which possess comparable values, reveal a similar evolution trend with the stoichiometry $x$ in $Li_xNi_{0.5}Mn_{0.3}Co_{0.2}O_2$. Meanwhile, the fitted linear relationship between $t_{ct}$ and $t_{0.9}$ with $R^2$=0.999 (Figure 3f) indicates that $t_{ct}$ and $t_{0.9}$ both measure the time scale of the same kinetic process, namely the charge transfer. Such correlation was also explored in Li-Li symmetric cells and SLPE cells (Figure S6a,b). The high consistency of the results from the time-domain method and the frequency-domain method proves that it is feasible to derive $\eta_{ct}$ and $\eta_s$ from the potential-time plot. It should be noted that the proposed overpotential method can be used to characterize the dynamic polarization behaviors, which are far beyond the capability of the static-state based EIS measurements.

Based on the separated $\eta_{ct}$, charge transfer kinetics at the particle-electrolyte interface can be precisely characterized. For a specific current density in the Tafel analysis (Figure 4a), the $\eta_{ct}$ derived from SLPE measurements is much smaller than that from the regular CE testing method [48,49], in which $\eta_{ct}$ is approximated by the total overpotential extracted from charging/discharging tests. The overestimation of $\eta_{ct}$ by CE measurements can cause great deviations of the subsequently calculated kinetic parameters. For instance, the estimated reaction constant by the Butler-Volmer (B-V) model from the CE measurements (Equation S22 and S23) is almost one order of magnitude lower than that from the SLPE measurements ($5\times10^{-12}$ vs. $3\times10^{-11}$ $m^{2.5}$ $mol^{-0.5}$ $s^{-1}$). In addition, the separated $\eta_{ct}$ shows a discrepancy from the classical Butler-Volmer model when the overpotential is over 100 mV. Even though similar phenomena, which are termed as curved Tafel plots, have been argued to be the effect of other overpotentials ($\eta_s$, $\eta_e$, and $\eta_l$) in the charging/discharging testing [49-51], our measurements directly demonstrate that the only $\eta_{ct}$ can still lead to the curved Tafel plots. Figure 4b also reveals that the B-V model deviates from the practically concentration-dependent Li intercalation kinetics, in which the overpotential calculated from the B-V model exhibits a hysteretic rise relative to the measured values. Such results may be due to the surface crowding effects of intercalation compounds [39,52], which are not taken into account in the B-V model.

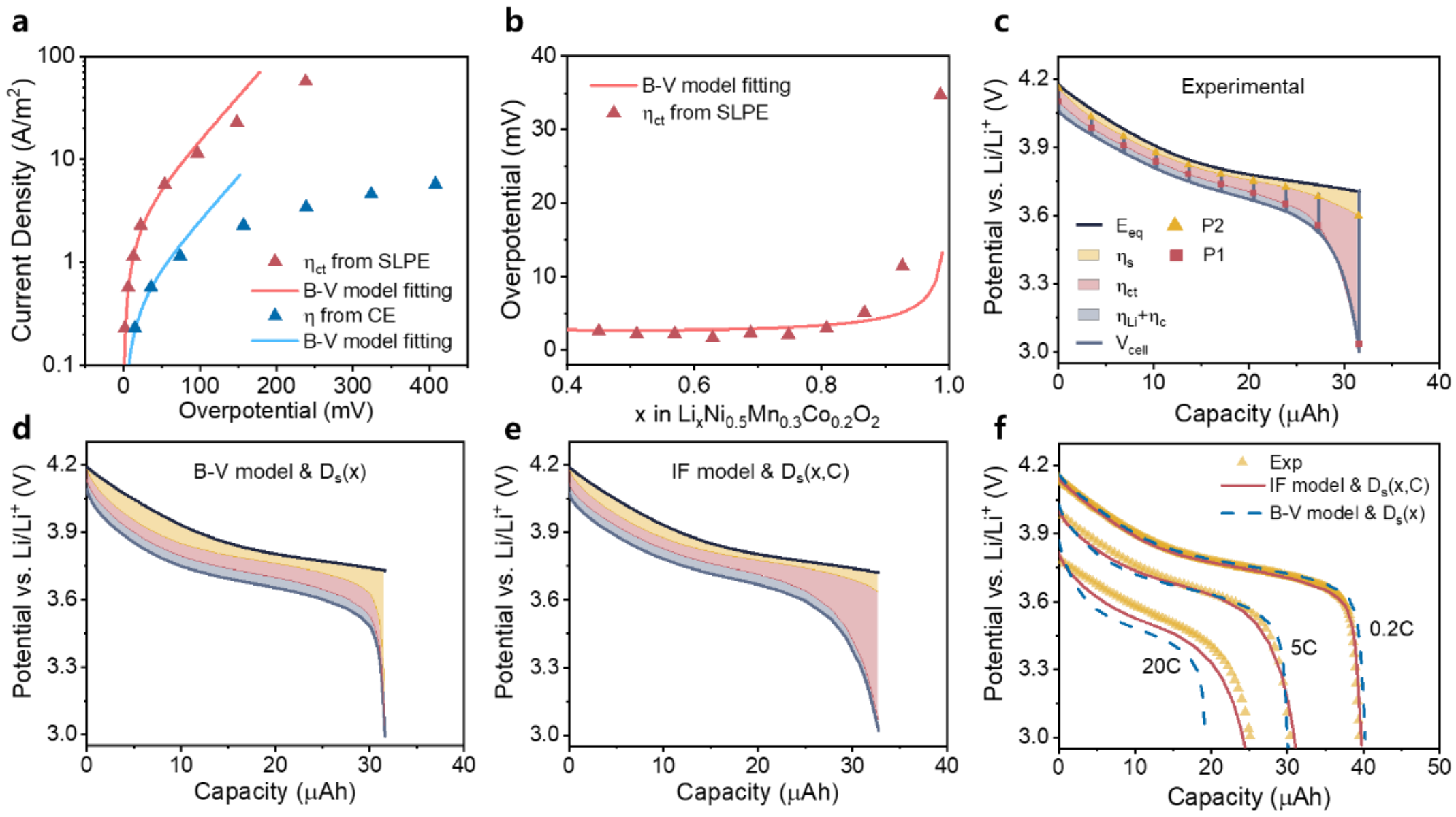

**Figure 4 Dynamic evolution of overpotentials of SLPES.** a,b, Comparison of experimentally measured overpotentials (discrete symbols) of charge transfer at various current densities (a) and Li stoichiometry of NMC532 (b) in a coin-cell configuration with the B-V model based prediction (solid lines). c, Measured overpotential components at various stages during a 5C discharge. d,e, Simulated overpotential evolution during a 5C discharge based on (d) the Butler-Volmer model with a rate-independent diffusivity (B-V model & $D_s(x)$) and (e) a interpolation function model with a concentration- and rate-dependent diffusivity (IF model & $D_s(x, \mathrm{C})$). f, Comparison of the electrochemical simulations (lines) and experimental discharging curves (dots) of SLPE cells at various C-rates.

The time-resolved potential measurements enable accessing dynamic evolution of overpotentials in charge/discharge operating, which cannot be realized by the static-state EIS measurements. Figure 4c reveals the overpotential components evolution captured by the current interruption method (Figure 3b) with a 1 s current interruption during the discharge of an SLPE cell at 5C rate. The measured $\eta_s$, $\eta_{ct}$, and $\eta_{Li+}\eta_c$ evolve as the discharge proceeds, and the charge transfer overpotential $\eta_{ct}$ takes the dominant role gradually, which are in distinct contrast to the simulated results from the B-V model and the Fick diffusion model with a concentration-dependent diffusivity (B-V model & $D_s(x)$, Figure 4d) that are widely used in LIB simulations. As the B-V model & $D_s(x)$ simulation results with the solid diffusion overpotential $\eta_s$ rather than $\eta_{ct}$ as the dominant overpotential component would lead to the explanation of kinetic mechanism inconsistent with experimental observations, we implemented simulations based on a constructed interpolation function (IF) model (Equation S24-S28 and Figure S10c) together with a concentration- and rate-dependent diffusivity ($D_s(x, \mathrm{C})$, Figure S11), as shown in Figure 4e (IF model & $D_s(x, \mathrm{C})$). The comparisons of the experimental measured discharging curves of SLPE cells at various C-rates with the simulated results from both the B-V model & $D_s(x)$ and the IF model & $D_s(x, \mathrm{C})$ are made in Figure 4f. It is clear that, based on SLPE measurements, the overpotential decomposition enabled IF model & $D_s(x, \mathrm{C})$ (Equation S9-S25) can perfectly reproduce the potential profiles, even in extreme polarization conditions (20C). On the contrary, the results from the B-V model & $D_s(x)$ show obvious departures from the experimental data when the rate is over 5C (Figure S12). Both our experiments and simulations indicate that the reaction-limited kinetics mechanism (Figure 4e) is more rational than the diffusion-limited mechanism shown in Figure 4d. Such conclusion is also supported by significant improvement in rate capability of thin NMC333 electrodes after altering the salt ingredient of electrolyte [43]. Moreover, our overpotential decomposition exhibits a significant perspective in electrode kinetic investigation.

## Overpotential decompositions of CE cells

Our proposed method is further extended herein for the investigation of overpotential components of CEs. The overpotential components of CEs vary with the running time (or DoD), operating conditions (e.g. C-rate), and electrode/cell designs (e.g. electrode thickness and porosity), leading to different dominating processes limiting the cell voltage and capacity. According to Newman's porous electrode theory [53,54], a continuum medium model (Equation S30-S52) is developed to calculate the overpotential components of CEs based on decoupled overpotential measurements from SLPEs. For a specific electrode design (Design 1 in Table S2), the dominant components of overpotentials change with C-rates (Figure 5a-c). At a low rate of 0.5C, the dramatic growth of the solid concentration overpotential ($\eta_s$) at the end of discharge leads to the cell voltage $V_{cell}$ hitting the cut-off line earlier relative to $E_{eq}$. For the case of 1C, the charge transfer overpotential ($\eta_{ct}$) also takes an important role in limiting the cell capacity, and the overpotential proportions of liquid ion transport, including the electrode part ($\eta_l$) and the separator part ($\eta_{sep}$), become more significant. When the discharge rate increases to 3C, $\eta_l$ and $\eta_{sep}$ increase sharply, causing the voltage plateau drops to only 3.2 V and the specific capacity declines to 63 mAh/g (only about a half of that at 1C). Surprisingly, $Li^+$ transport in the porous separator becomes the uppermost factor limiting fast discharging, according to the largest value of $\eta_{sep}$ at 3C. This may be the reason that the salt concentrates in the separator owing to the sharply increased $Li^+$ flux from the Li meta plane at high C-rates (Figure S14c,d), resulting in dramatically reducing of the electrolyte conductivity due to electrostatic interaction in concentrated solution [55,56]. Our result implies that the role of separators has been overlooked in high-power applications, since the limiting process at high C-rates was widely believed to be the electrolyte transport in the electrode pores [57,58].

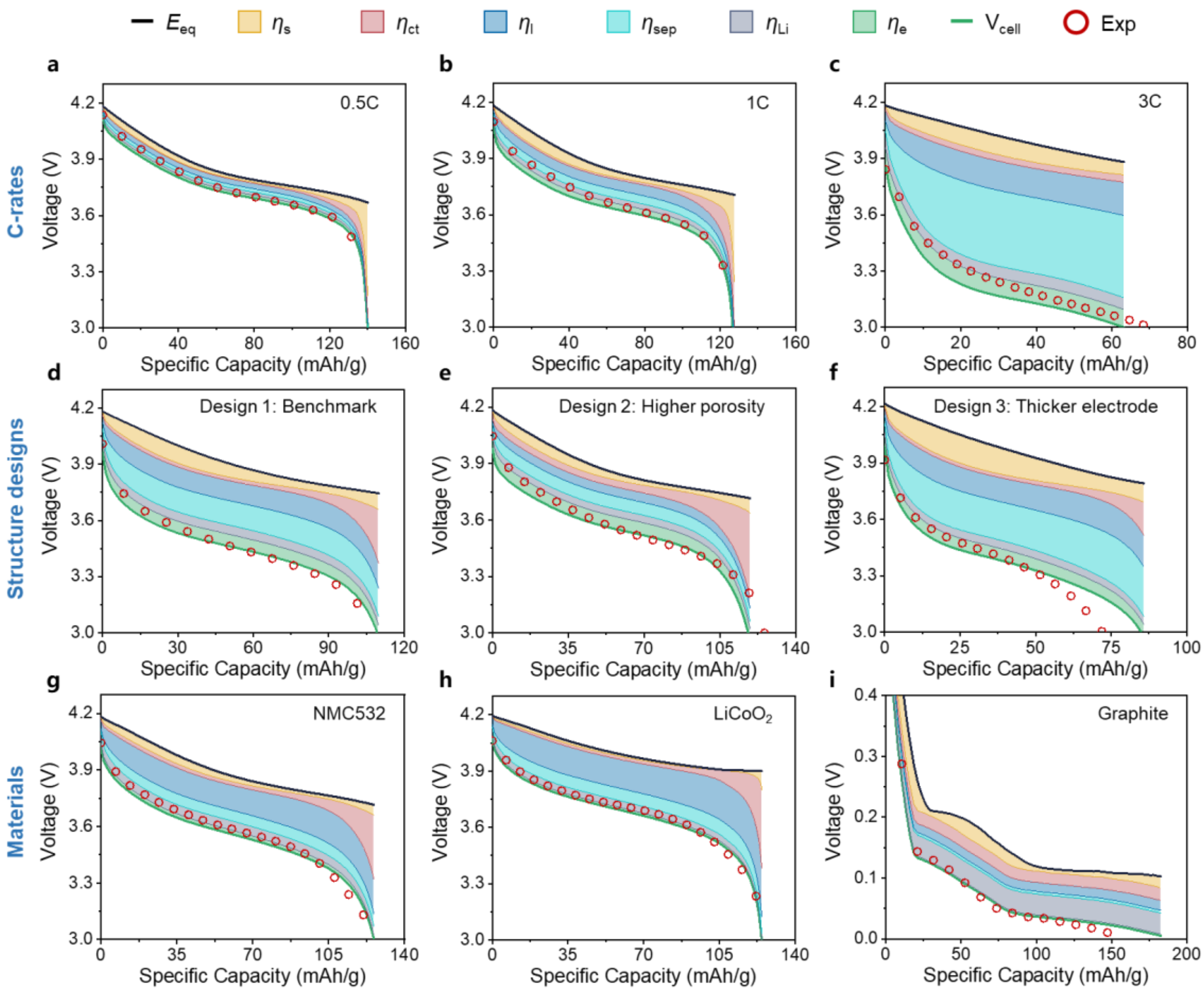

**Figure 5 Overpotential decompositions of CE cells.** a-c, Overpotential components of Design 1 at 0.5C (a), 1C (b), and 3C (c) rates, respectively. d-f, Overpotential components of Design 1 (e), Design 2 (f), and Design 3 (g) at 2C rate, respectively. g-i, Overpotential components of CE cells with NMC532 (g), $LiCoO_2$ (h), and graphite (i), respectively.

Based on the overpotential decomposition, the limiting processes in CE cells can be identified. As manifested in Figure 5d-f, different CEs with various electrode designs are tested at the same C-rate (2C). In the control design (Design 1), $\eta_{ct}$, $\eta_l$, and $\eta_{sep}$ together compel the discharge to end at a specific capacity of 107 mAh/g (Figure 5d). With the porosity increasing from 0.39 (Figure 5d) to 0.55 (Figure 5e), the proportion of $\eta_{ct}$ becomes the largest in the total overpotential for Design 2, indicating the charge transfer is the primary limiting process. Both $\eta_l$ and $\eta_{sep}$ decrease due to the large porous space in the electrode accessible to the electrolyte for ion transport, which improves the voltage plateau. In addition, the increment of electrode thickness from 120 μm (Design 1) to 145 μm (Design 3) can cause an even more significant $\eta_{sep}$. As the $Li^+$ flux from Li metal increases, the high areal loading can also lead to the aggregation of salt in the separator pores (Figure S15b) similar to the case of Figure 5c, seriously limiting the electrode capacity at high C-rates. Moreover, the overpotential decompositions of various material systems, including NMC532, $LiCoO_2$, and graphite, are demonstrated in Figure 5g-i. The accurate overpotential decomposition of CEs enables reliable predictions of voltage evolutions of CEs with different materials and designs (Figure S16) in a wide range of C-rates (0.1-5C), which was hardly achieved in previous studies [16,59,60].

With the identification of limiting processes in CE cells via overpotential decomposition, the optimization of electrochemical kinetics can be further conducted. Referring to Design 1 in Figure 5c, mitigating the concentration polarization of the separator should be the key to improving the rate performance. By reducing the tortuosity of NMC532 electrodes to enhance ion transport in the electrode pores, alleviating the aggregation of ions in the separator pores to some extent (Figure S15a), the predicted discharge curve and overpotential contributions at 3C rate show that reducing the tortuosity from 2.8 to 1.6 improves the specific capacity from 67 mAh/g to 97 mAh/g (Figure 6a). According to the previous experiments that optimized the electrode pore structure by constructing vertically-oriented pores to enhance the $Li^+$ transport kinetics [61,62], our fabricated oriented-pore electrodes with the effective tortuosity of 1.3 and 2.1 confirm the improvement of rate performance, showing an increase in the specific capacities to 112 mAh/g and 94 mAh/g at a 3C rate, respectively (Figure 6b,c). For Design 3 with a thicker electrode, the separator needs to be optimized with lower MacMullin number $N_M$ (defined as tortuosity divided by porosity) to boost ion transport in separator pores (Figure S15b). As predicted by our model, an attractive increment in specific capacity (around 105 mAh/g at 3C rate) may be achieved when reducing $N_M$ from 10.4 to 4.4 (Figure 6d). The corresponding experiment results show that the cell with $N_M$=4.9 and 4.4 (Table S3) display much better rate performance than that with $N_M$=10.4 (Figure 6e, f). The cells with $N_M$=4.4 possess about 3-fold higher capacity (95 mAh/g) compared to that with $N_M$=10.4 (27 mAh/g) at 3C rate, as predicted by our model calculations. Therefore, our overpotential decompositions can provide valuable guidance for the rational design of electrode/cell structures.

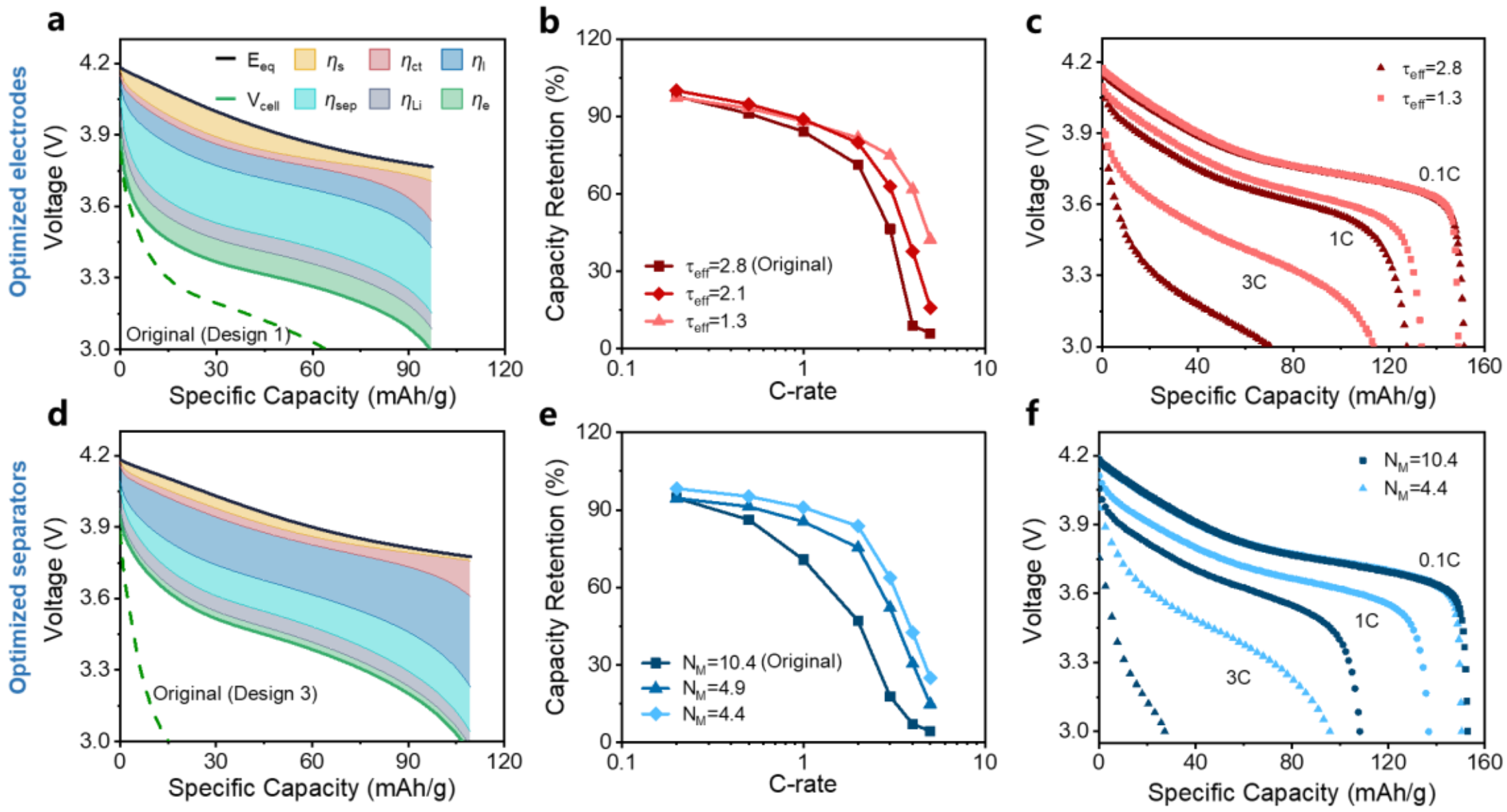

**Figure 6 Overpotential decomposition based design and optimization of electrodes/cells.** a, Predicted overpotential components (solid lines) at 3C rate via the optimization of electrode pore structure in Design 1 (dashed line). b, Experimental validation of the rate capability improvement of CEs with different effective tortuosity $\tau_{eff}$, i.e. 2.8 (corresponding to Design 1), 2.1, and 1.3. c, Discharge voltage curves of the CEs with $\tau_{eff}$=2.8 and $\tau_{eff}$=1.3 at 0.1C, 1C, and 3C. d, Predicted overpotential components (solid lines) at 3C rate via the optimization of separator structure in Design 3 (dashed line). e, Experimental validation of the rate capability improvement of CEs with different MacMullin number $N_M$, i.e. 10.4 (corresponding to Design 3), 4.9, and 4.4. f, Discharge voltage curves of the CEs with $N_M$=10.4 and $N_M$=4.4 at 0.1C, 1C, and 3C.

## Conclusion

Overpotentials of LIB electrodes were decoupled via a well-defined SLPE structure combined with time-resolved potential measurements, explicitly revealing the evolution of kinetic processes that are dependent on operating conditions and structure designs. The separated charge transfer overpotential of active particles shows that the classical Butler-Volmer model may be not reliable enough at a high state of lithiation or large current density, leading to incorrect overpotential components and subsequent misunderstanding of the corresponding kinetic limitation mechanisms. By constructing an interpolation function model from SLPE based potential measurements, perfect prediction of the discharging profiles can be achieved, even in extreme polarization conditions (20C). Moreover, the applicability of the overpotential decomposition method in different electrode/cell structures and material systems is further demonstrated, uncovering the limiting processes of battery rate performance, based on which the optimization of electrochemical kinetics can be conducted. Thereby, our study not only shades light on decoupling complex kinetics in electrochemical systems, but also provides vitally significant guidance for the design of high-performance batteries.

## Methods

**Electrode fabrication and characterization.** For the SLPEs, a suspension of NMC532 (MTI), carbon black (Super-P, MTI), polyvinylidene fluoride (PVDF, Arkema), and N-methyl-2-pyrrolidone (NMP, Sinopharm Chemical Reagent Inc.) at a mass ratio of 3:0.5:1:20 was first mixed at a rate of 500 r/min for 6.5 h to be dispersed sufficiently. The prepared suspension was then coated on an aluminum foil by a blade coater with a gap of 40 μm. Subsequently, the wet electrode was dried at 100 °C by infrared radiation for 3 h. Different mass ratios of the suspension components and coating gaps were also tried for comparison. For the CEs, the fabricating process was almost the same as that of the SLPEs. The component ratios were NMC/Super-P/PVDF/NMP 20:1:1:20 in wt/wt/wt/wt and the coating gaps were 400-750 μm. The CEs were calendered to a certain porosity (about 0.4) by two steel rollers of a roller press. Before assembling these electrodes into cells, a baking process was conducted at 80 °C in a vacuum.

The particle morphology and distribution of the SLPEs were observed by field-emission scanning electron microscopy (Sirion 200, FEI) in high-vacuum mode with an acceleration voltage of 5 kV. The electrode weights were measured by using a microbalance (METTLER TOLEDO, ME55), and the electrode thicknesses were determined by a digital micrometer. The size distribution of NMC532 particles was acquired by a laser particle size analyzer (Mastersizer 3000, Malvern), and the packing density of active particles was directly measured from the SEM images.

**Cell assembly and measurements.** Some SLPEs and all the CEs were assembled into 2032-type coin cells with a Li-metal counter electrode and a commercial separator (Celgard) soaked in an electrolyte solution (DoDoChem). The separators used in the SLPE cells were Celgard 2400, while in the CE cells were Celgard 2325, 2400, or 2500 for comparison. The ingredients of the electrolyte were 1M $LiPF_6$ and EC/EMC/DMC at a volume ratio of 1:1:1 with 1.0% VC, and the Li salt concentration for some SLPE cells was changed to 0.5M or 1.5M for comparison. Other SLPEs were assembled in three-electrode cells (MTI) with a Li metal foil as the counter electrode and a Li metal ring as the reference electrode (Celgard 2400 separator and 1M $LiPF_6$ in the electrolyte). In Li-Li symmetric cells, Celgard 2400/2500 and the electrolyte with 1M $LiPF_6$ were used. These commercial separators were also assembled into the coin cells with the electrolyte solution for conductivity testing.

Charge/discharge tests were performed on a battery cycler (LANHE, M340A). The voltage windows of the SLPE cells were 3-4.2 V and 3-4.5 V, while that of the CEs was 3-4.2 V. In the formation step and the specific capacity test, a constant-current constant-voltage (CC-CV) charging and CC discharging (CC-CV/CC) protocol was applied, in which the cells were first charged to 4.2 V or 4.5 V at a rate of 0.1C and then charged at the fixed voltage until the current decreased to 0.05C in the charging process. The equilibrium potential was approximated by the voltage measured at 0.1C rate. The rate-discharge performance was measured at 0.2C, 0.5C, and 1-5C for the CE cells and 0.2C, 0.5C, 1C, 2C, 5C, 10C, 20C, and 50C for the SLPE cells by CC-CV/CC protocols as well (only CC discharging for the Li-Li cells). The cycling tests adopted CC/CC protocols, in which the SLPE cells were cycled at 2C and the CE cells were at 0.5C. The three-electrode cells were first charged to 4.2 V at 0.1C with a CC-CV protocol and then repeatedly rested for 1h and discharged at C/8 for 1 h until the potential dropped to 3 V. The sampling frequency was set as 1 kHz during the initial 10 s of every discharge stages. After each discharging and resting stage, the EIS of the SLPEs vs. the reference electrodes ($Li/Li^+$) was measured over the frequency range from 0.01 Hz to 1 MHz on an electrochemical workstation (Gamry, 1010E). The EIS of the SLPE cells, CE cells, Li-Li cells, and separators were also measured in the same frequency range. The CV tests of the SLPE cells were conducted between 3 V and 4.2 V at a scan rate of 0.1 mV/s. The galvanostatic intermittent titration technique (GITT) tests were performed on the SLPE cells. For each titration, a current pulse was applied at 0.1C rate for 30 min, followed by a 1 h relaxation period. All the measurements above were conducted at room temperature (25 °C) in a temperature-controlled room.

**Electrochemical modeling and simulations.** In the electrochemical model of the SLPE cells, a representative volume element (RVE) was modeled to represent the entire electrode sheet of SLPEs. We generated the geometrical parameters (position, radii) of the NMC532 particles in the RVE by random function in Matlab, and the areal capacity and the average diameter are consistent with the experimental cells. The SLPE cell model was assumed to be isothermal, and the current collector was assumed to be isopotential. The $Li^+$ diffusion process within active particles was described by Fick's second law and the mass/charge transport in the electrolyte domain was modeled by concentrated solution theory. The interpolation function model was used for the charge transfer kinetics of NMC532 particles, while we retained the Bulter-Volmer model for Li metal anodes. The detailed mathematic description and model parameters are given in Equation S4-S20 and Table S1, respectively.

The CE cells were modeled by Newman's method in which the porous electrodes were treated as a continuum medium with the volume averaging method. The CE cell model only considers the kinetic processes in the electrode thickness direction by assuming a uniform electrode plane. In the model, the porous electrode and the electrolyte are superimposed in space, and the electrode porosity and tortuosity are introduced to consider the effects of porous media on transport physics. The detailed mathematic description and model parameters are given in Equation S30-S52 and Table S2.

## Conflict of Interest

The authors declare no conflict of interest.

## Acknowledgment

This work was supported by the National Natural Science Foundation of China (Grant No. 52175317, No. 12172143) and the Fundamental Research Funds for the Central

Universities (No. 2021yjsCXCY023).

# Supplementary Information

# Overpotential Decomposition Enabled Decoupling of Complex Kinetic Processes in Battery Electrodes

Ruoyu Xiong[1], Yue Yu[1], Shuyi Chen[1], Maoyuan Li[1], Longhui Li[1], Mengyuan Zhou[1], Wen Zhang[2], Bo yan[3], Dequn Li[1], Hui Yang[2,*], Yun Zhang[1,*], Huamin Zhou[1,*]

## Calculation of packing density $n_A$

The total mass of active particles in the suspension can be expressed by:

$$m_{ap} = \sum_i V_{ap,i}\rho_{ap} = \frac{4\pi N \rho_{ap} \overline{R_{ap}}^3}{3} \quad \text{(S1)}$$

where $m_{ap}$ and $\rho_{ap}$ denote the total mass and the specific mass of active particles in the suspension, respectively, $V_{ap,i}$ the volume of the $i$-th active particle, $N$ the total number of active particles in the suspension, $R_{ap}$ the equivalent radii of active particles, $V$ the volume of the suspension. We define the spatial distribution density (the number of active particles per unit volume) in the suspension as:

$$n_V = \frac{N}{V} \quad \text{(S2)}$$

where $V$ is the volume of the suspension. The relation between the packing density $n_A$ and $n_V$ is

$$n_{\mathrm{A}} = n_V l \quad \text{(S3)}$$

where $l$ is the coating gap. Substituting Equation S1 and Equation S2 into Equation S3, then we have Equation 1.

## SEM characterization of SLPEs

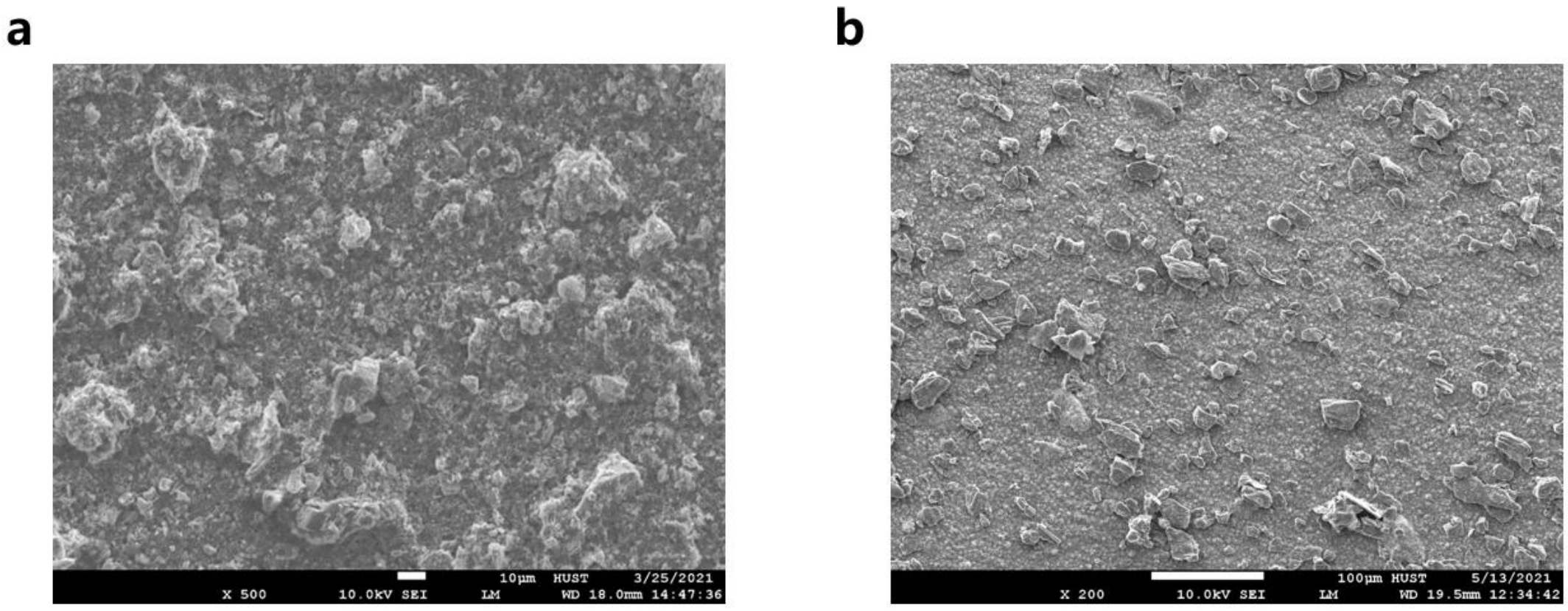

**Figure S1** SEM images of $LiCoO_2$ SLPEs (a) and graphite SLPEs (b).

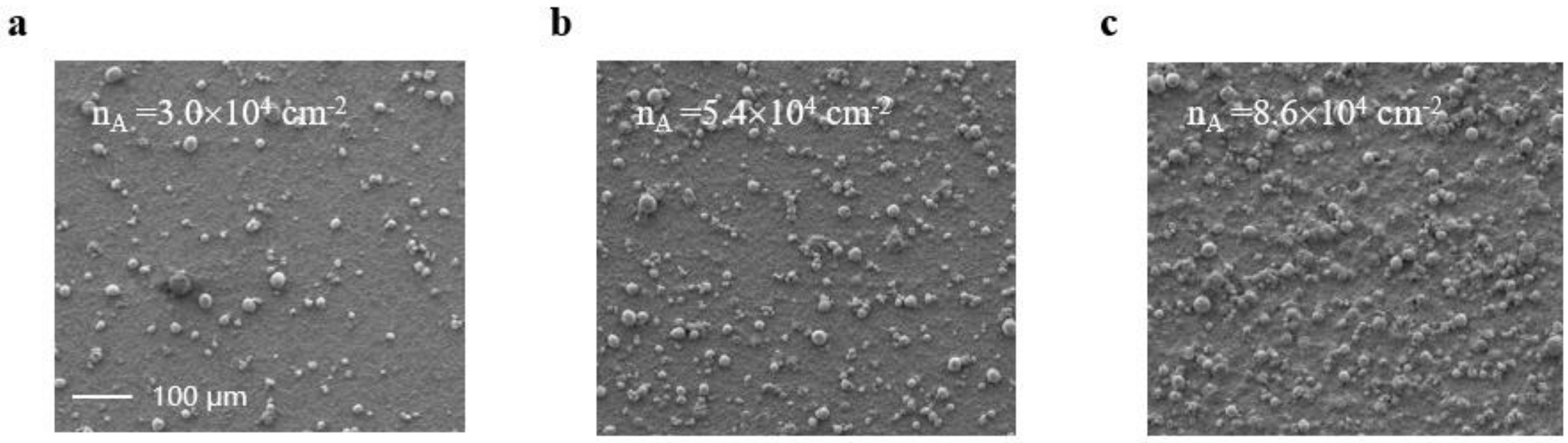

**Figure S2** SEM images of NMC532 SLPEs with different packing densities.

## Electrochemical characterizations of SLPEs

In the voltage windows of 3-4.2 V and 3-4.5 V, the SLPEs averagely released 151 mAh/g and 185 mAh/g, respectively (Figure S3a). Comparing the cases of the two packing densities ($n_A$ = 1.7×10$^5$ and 3.7×10$^5$ cm$^{-2}$), there is no apparent difference in specific capacity between them. Typical redox peaks can be seen in the cyclic voltammetry profile (Figure S3b), indicating that no other reaction but Li insertion/extraction took place in the SLPEs. In addition, the SLPE cells display much more excellent rate capability than the CE cells with an areal capacity of 3.8 mAh/cm$^2$, as shown in Figure S3c,d. The capacity retention of the SLPE cells is around 50% even at 50C, while that of the CE cells drops to a poor level when the C-rate exceeded 3C. The results are amazing but reasonable since the SLPEs do not have long transport paths for electrons or ions, which are generally the rate-limiting processes at high C-rates in a CE. According to the cycling performance illustrated in Figure S3e, the SLPE cells show a slower rate of capacity decline even at a much higher C-rate (charging/discharging at 2C) than that for the CE cells (charging/discharging at 0.5C). The SLPE cells also exhibit better durability while the CE cells fail after less than 90 cycles.

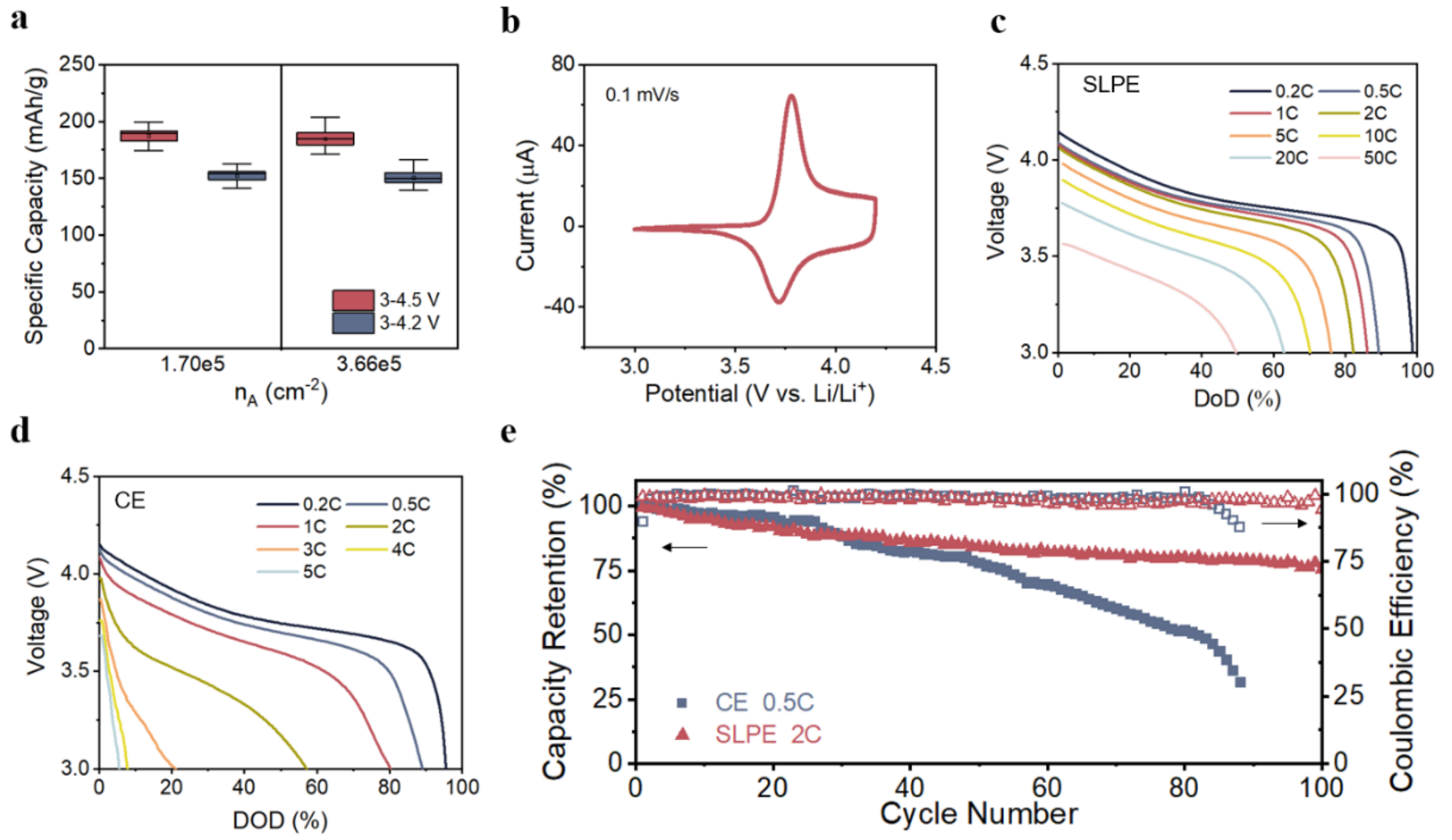

**Figure S3** a, Specific capacities of the SLPEs with $n_A = 1.7\times10^5$ cm$^{-2}$ (left) and $n_A = 3.7\times10^5$ cm$^{-2}$ (right), respectively. b, Cyclic voltammetry curve of SLPEs with a scan rate of 0.1 mV/s. c,d, Rate-discharging curves of SLPEs (c) vs. CEs (d). e, Cycling performance of SLPEs vs. CEs with CC/CC procedures (0.5C for CE cells and 2C for SLPE cells).

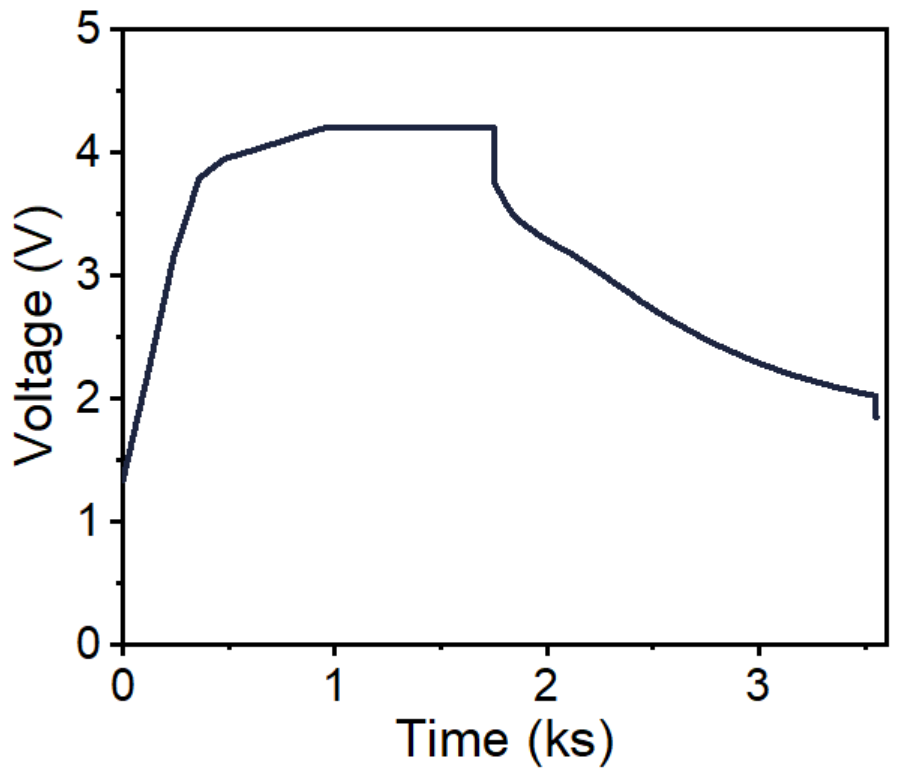

**Figure S4** Voltage curves of SLPEs without CB (0 wt%) at the first cycle.

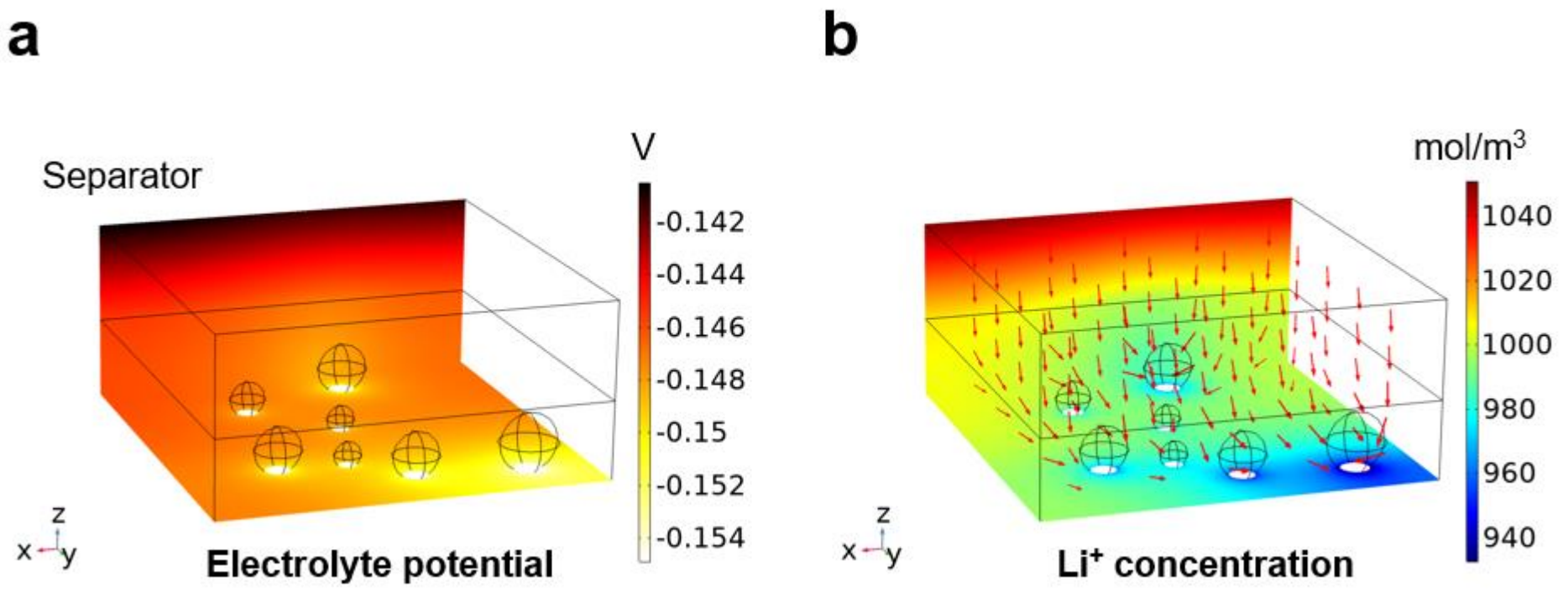

**Figure S5** Electrolyte potential (a) and Li$^+$ concentration (b) distributions of the SLPE cell with 1M of Li salt concentration and $7\times10^4$ cm$^{-2}$ of packing density at the end of discharge (50C) from electrochemical simulations.

## Electrochemical model of SLPE cells

**Model assumptions.** A representative volume element (RVE) is modeled to represent the entire electrode sheet of SLPEs. We generate the geometrical parameters (position, radii) of the NMC532 particles in the RVE by random function in Matlab, and the areal capacity and the average diameter are consistent with the experimental cells (Table S1). The SLPE cell model is assumed to be isothermal, and the current collector is assumed to be isopotential. More assumptions are pointed out near the specific mathematical equations below.

**Governing equations.** In the active particles, the diffusion process of Li (polarons) is described by Fick's second law

$$\frac{\partial c_s}{\partial t} = \nabla \cdot \left( D_s \nabla c_s \right) \tag{S4}$$

where $c_{\mathrm{s}}$ is the normalized Li concentration in the active particles, $D_s$ is the apparent diffusion coefficient. The electrostatic field is described by Ohm's law

$$0 = \nabla \cdot \left( -\sigma \nabla \phi_s \right) \tag{S5}$$

where $\phi_s$ denotes the electrode potential, σ the conductivity of the composite electrode.

Based on the concentrated solution theory [1,2], the transport process of $Li^+$ ions in the electrolyte is governed by

$$\frac{\partial c_l}{\partial t} = \nabla \cdot \left( D_l \nabla c_l \right) - \nabla \frac{t_+ \mathbf{i}_l}{F} \tag{S6}$$

where $c_l$ is the normalized $Li^+$ concentration of the electrolyte solution, $D_l$ the diffusion coefficient, $t_+$ the $Li^+$ transference number, $\mathbf{i}_l$ the current density in the liquid phase, and $F$ the Faraday constant. The charge transport in the electrolyte solution can be expressed as

$$0 = \nabla \cdot \left( -\kappa \nabla \phi_l + \frac{2RT\kappa \left(1 - t_+^0\right)}{F} \left( 1 + \frac{\partial \ln f_\pm}{\partial \ln c_l} \right) \nabla \ln c_l \right) \tag{S7}$$

where $\phi_l$ is the electrostatic potential of the electrolyte solution, $\kappa$ the electrolyte conductivity, $f_\pm$ the activity coefficient of the electrolyte, and $R$, $T$ are the universal gas constant and the testing temperature, respectively. In the separator domain, we consider the effects of pore structure on transport kinetics by modifying $D_l$ and $\kappa$

$$D_{l,eff} = \frac{\varepsilon_{sep}}{\tau_{sep}} D_l , \quad \kappa_{eff} = \frac{\varepsilon_{sep}}{\tau_{sep}} \kappa \tag{S8}$$

where $\varepsilon_{sep}$ and $\tau_{sep}$ are the porosity and tortuosity of the separator, respectively.

The reaction current density $i$ at the particle-electrolyte interface is given by Equation S28 combined with the concentration dependence of $Li^+$ in the electrolyte.

$$i = \left(c_l / c_{l,0}\right)^{0.5} h\left(\eta_{NMC,ct} \Big/ \sqrt{g\left(c_{s,surf} / c_{s,\max}\right)}\right) \tag{S9}$$

where $c_{s,\max}$ and $c_{s,surf}$ denote the maximum and surface Li concentration of active particles, respectively, and $c_{l,0}$ is the initial $Li^+$ concentration of the electrolyte. The overpotential $\eta_{NMC,ct}$ is defined as:

$$\eta_{NMC,ct} = \phi_s - \phi_l - E_{eq,surf} \tag{S10}$$

where $E_{eq,surf}$ denotes the thermodynamic equilibrium potential of the intercalation compounds at the particle surface. The contact resistance between the active particle and the current collector is considered in the cell voltage calculation

$$V_{cell} = \phi_s - IR_c \tag{S11}$$

The reaction kinetics of the Li metal electrode is expressed as

$$i_{Li} = k_{Li} \left( \frac{c_l}{c_{l,0}} \right)^{0.5} \left[ \exp\left( \frac{F\eta_{Li,ct}}{2RT} \right) - \exp\left( -\frac{F\eta_{Li,ct}}{2RT} \right) \right] \tag{S12}$$

The SEI film contributes a film resistance $R_{sei}$ at the Li metal-electrolyte interface, hence the interfacial overpotential is

$$\eta_{Li,ct} = \phi_{Li} - \phi_l - IR_{sei} \tag{S13}$$

**Boundary conditions.** The boundary condition of the solid diffusion in the active particles is

$$\left(-D_s \nabla c_s\right) \cdot \mathbf{n} = \frac{i}{F} \tag{S14}$$

where **n** is the unit normal vector of the electrode-electrolyte boundary. The boundary condition of the electrostatic field at the particle-electrolyte interface is

$$\left(-\sigma\nabla\phi_s\right)\cdot\mathbf{n}=i \tag{S15}$$

and at the particle-current collector interface is

$$\int\left(-\sigma\nabla\phi_s\right)\cdot\mathbf{n}=I \tag{S16}$$

The boundary conditions of the species transport in the electrolyte solution are

$$\left(-D_l\nabla c_l\right)\cdot\mathbf{n}=\frac{i}{F},\quad \left(-D_l\nabla c_l\right)\cdot\mathbf{n}=\frac{i_{Li}}{F} \tag{S17}$$

for the particle-electrolyte interface and the Li metal-electrolyte interface, respectively. The boundary conditions of the charge transport are

$$\left(-\kappa\nabla\phi_l\right)\cdot\mathbf{n}=i,\quad \left(-\kappa\nabla\phi_l\right)\cdot\mathbf{n}=i_{Li} \tag{S18}$$

for the particle-electrolyte interface and the Li metal-electrolyte interface, respectively. The other boundaries of the electrolyte domain are set as the periodic boundary condition

$$D_l\nabla c_l\cdot\mathbf{n}=0,\quad \kappa\nabla\phi_l\cdot\mathbf{n}=0 \tag{S19}$$

The electrostatic potential of the Li counter electrode is set as zero potential reference,

$$\phi_{Li}=0 \tag{S20}$$

**Table S1**

Model parameters used in the electrochemical simulation of SLPE cells.

| **Parameters** | **NMC532 particles** | **Separator** | **Li metal** |
|---|---|---|---|
| Thickness (μm) | — | 25 [b] | — |
| Porosity | — | 0.41 [b] | — |
| Tortuosity | — | 2.0 [a] | — |
| Diameter of active particle (D50, μm) | 12 [b] | — | — |
| Electrode conductivity (S $m^{-1}$) | 100 [a] | — | — |
| Solid-phase diffusivity ($m^2$ $s^{-1}$) | Equation S29 [c] | — | — |
| Maximum solid-phase concentration (mol $m^{-3}$) | 48152 [a] | — | — |
| Initial $x$ | 0.47±0.01 [a] | — | — |
| Reaction rate constant ($m^{2.5}$ $mol^{-0.5}$ $s^{-1}$) | $3\times10^{-11}$ [a] | — | $3.7\times10^{-7}$ [a] |
| Initial electrolyte concentration (mol $m^{-3}$) | 1000 [b] | | |
| Liquid-phase diffusivity ($m^2$ $s^{-1}$) | $2.2\times10^{-10}$ [c] × Fitting function [d] | | |
| Electrolyte conductivity (S $m^{-1}$) | 0.724 [b] × Fitting function [d] | | |
| $Li^+$ transference number | 0.363 [d] | | |
| Thermodynamic factor | Fitting function [d] | | |
| Contact resistance (Ω) | 50 [a] | | |
| SEI film resistance (Ω) | 100 [a] | | |

[a] Measurements and calculations, [b] Information from the manufacturer, [c] Estimated by fitting, [d] [Valoen, L. O. and J. N. Reimers (2005). "Transport properties of LiPF6-based Li-ion battery electrolytes." Journal of The Electrochemical Society 152(5): A882-A891.]

## EIS and potential measurements of Li-Li symmetric cells

As shown in Figure S6a, there is only one arc in each Nyquist plot of the Li metal electrodes, which is commonly believed to be the overlapping regions of $Li^+$ diffusion in the solid electrolyte interface (SEI) film and deposition/dissolution reaction of metal Li [3]. Besides, the impedance first shows an increase and then holds steady (Figure S7), which most likely originates from the formation of SEI film. When given a step current stimulation of 8 μA (Figure S6b), it exhibits an instantaneous jump of potential (less than 2 ms) as soon as the current is loaded to the specified value (the abnormal point can be ignored as well). Comparing the real part of the arc with the effective resistance measured by the potential response, we also find a good consistency between the values from the frequency domain (288 Ω) and the time domain (278 Ω). A similar phenomenon can be observed in SLPE-Li cells. As illustrated in Figure S6c, the impedance regions of the Li metal anode and the current collector/particle interface are superposed, while they are distinct from the charge transfer region. The overpotentials of the two parts are also easy to be distinguished in the potential plot, and the equivalent resistances ($R_c$+$R_{Li}$ = 257 Ω and $R_{ct}$ = 232 Ω) are approximate to that in EIS (about 240 Ω and 260 Ω, respectively).

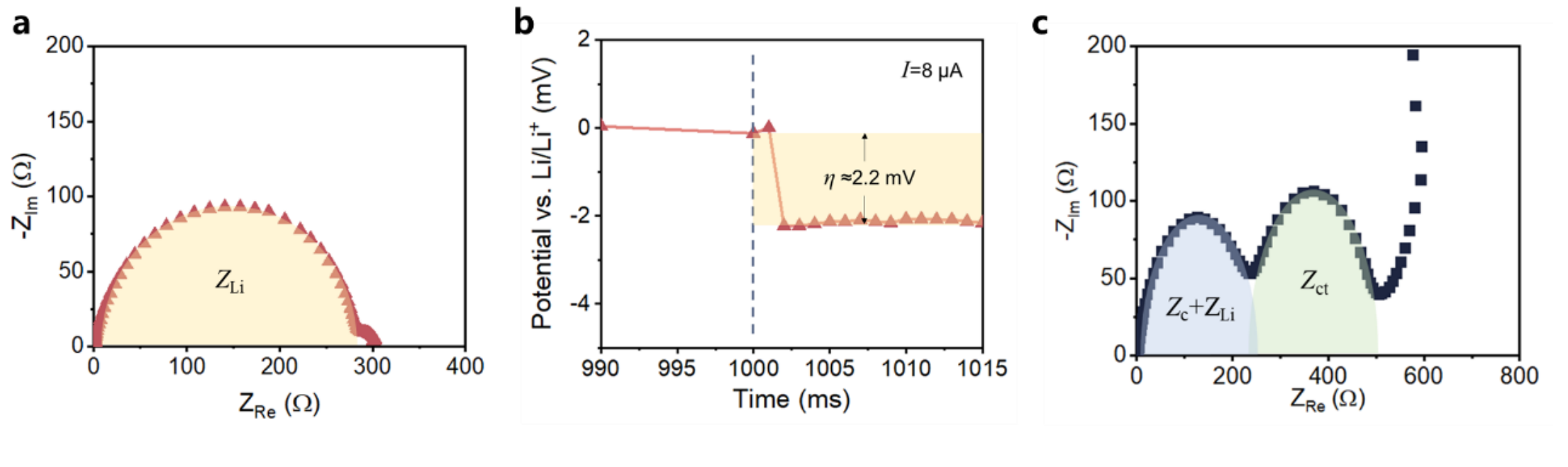

**Figure S6** a,b, EIS (a) and potential response (b) of Li metal electrodes (1/2 of Li-Li symmetric cells) to a current stimulation of 8 μA. c, EIS of SLPE cells.

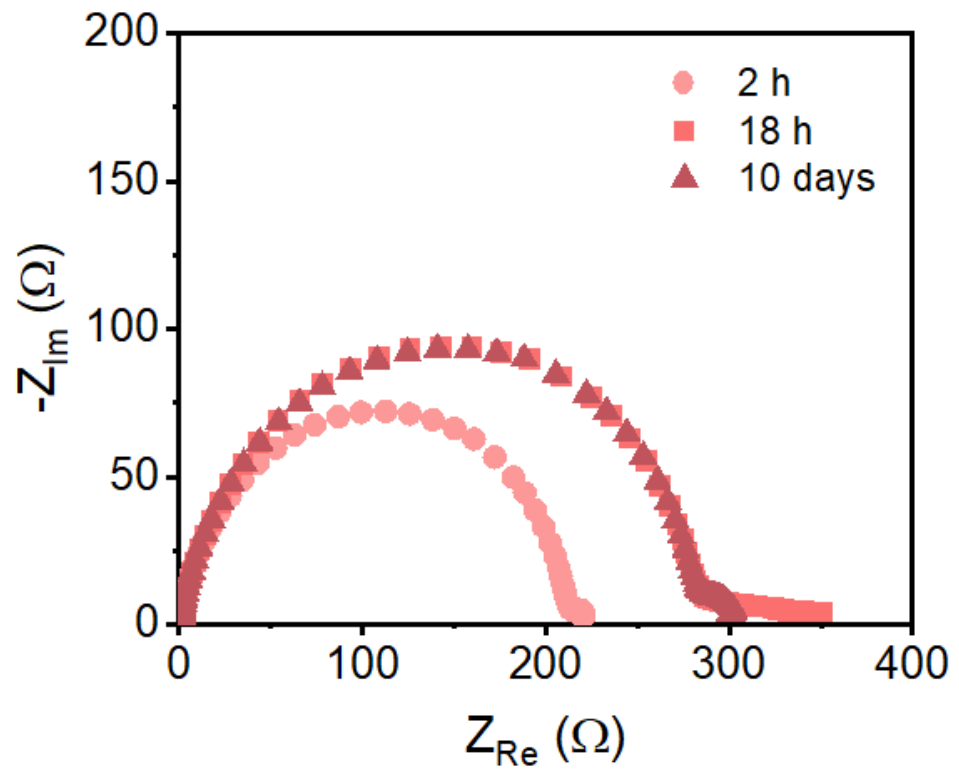

**Figure S7** EIS data of Li metal electrodes (1/2 of Li-Li symmetric cells) after resting 2 h, 18 h, and 10 days from just assembly.

## Piecewise discharge testing of SLPEs by three-electrode technology

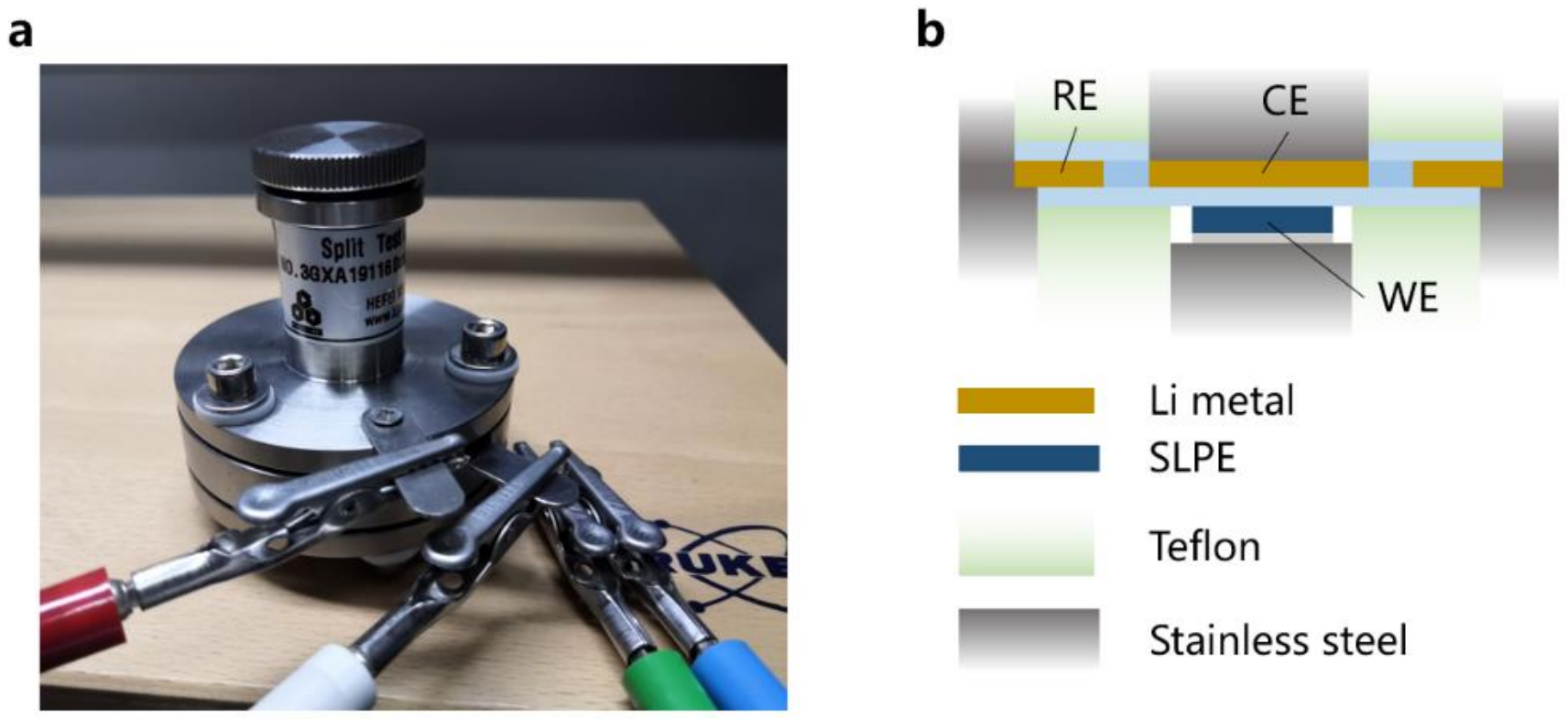

**Figure S8** Three-electrode configuration (a) and its electrochemical and mechanic structure (b).

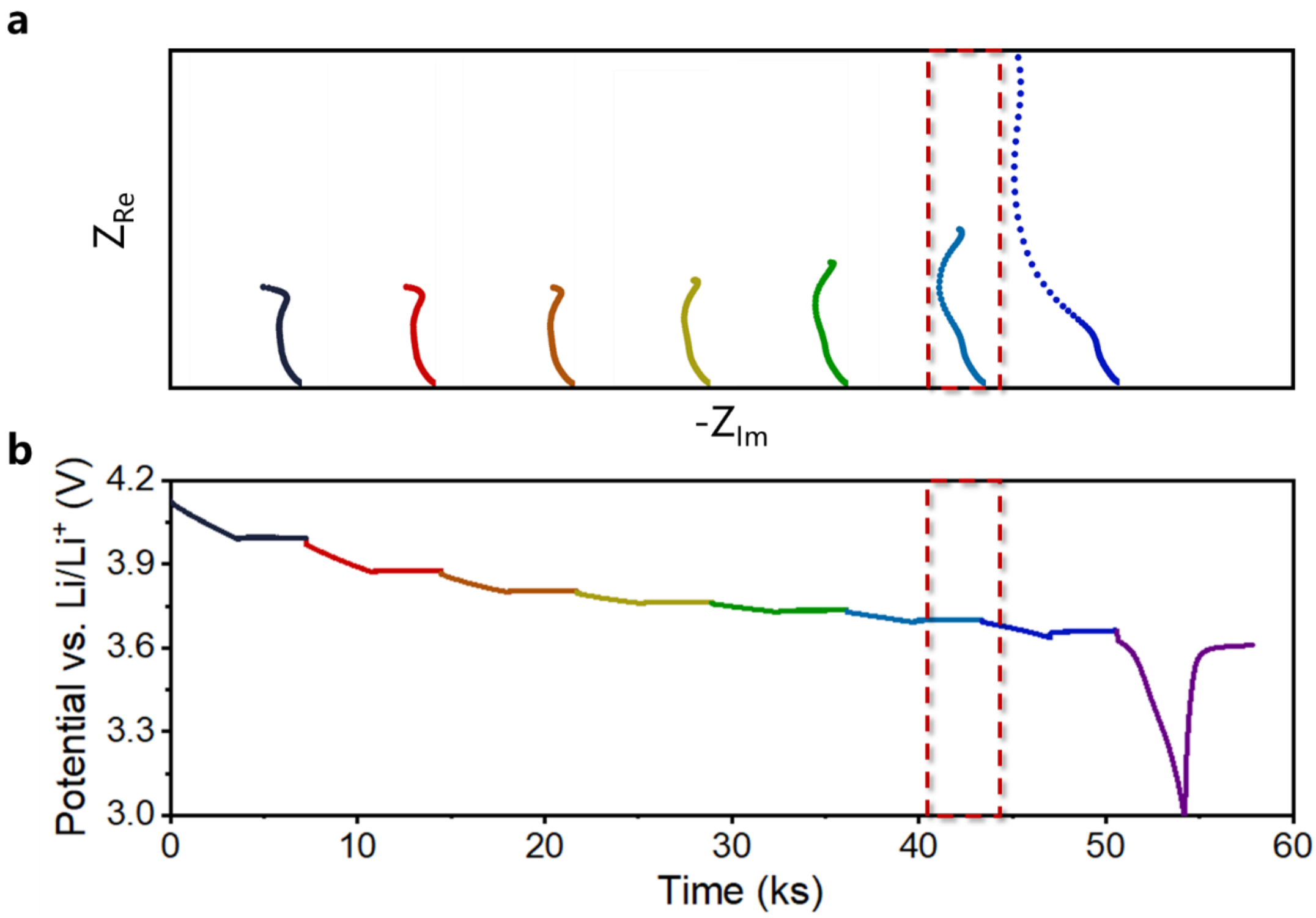

**Figure S9** Three-electrode analysis at both the time domain and the frequency domain. a, Potential measurements at varied DoD during a piecewise discharge, in which the discharge rate in every stage was 1/8C and each resting stage was set between every two discharge stages. b, EIS plots at varied DoD during the piecewise discharge. The EIS was collected after every discharging and resting stage of the piecewise discharge.

## Investigation of charge transfer kinetics via separated $\eta_{ct}$

As illustrated in Figure S10a the current-overpotential relation of $Li^+$ intercalation/deintercalation kinetics at NMC532 particle surfaces was obtained by symmetric pulsing charging/discharging at varied C-rates. The classical Butler-Volmer model, which is widely used to describe the reaction kinetics of LIB electrodes, is examined by the measured results. The general form of the Butler-Volmer equation is

$$i = i_0 \left[ \exp\left( \frac{\alpha F \eta_{ct}}{RT} \right) - \exp\left( -\frac{(1-\alpha) F \eta_{ct}}{RT} \right) \right] \quad \text{(S21)}$$

where $i$ and $\eta_{ct}$ denote the reaction current density and overpotential at the electrode-electrolyte interface, respectively, and $i_0$ is the exchange current density, $\alpha$ the transfer coefficient, $F$ the Faraday constant, $R$ the universal gas constant, and $T$ the testing temperature. It can be seen that the Butler-Volmer model can well describe the relationship between the current density (i.g. reaction rate) and overpotential. The phenomenon of curved Tafel plots mentioned in some literature is not significant in this small overpotential range [4-6]. The perfect symmetry of charging and discharging implies the transfer coefficient $\alpha$ = 0.5. The exchange current density $i_0$ in Equation S21 is defined as

$$i_0 = Fk \left( c_{s,\max} - c_{s,surf} \right)^{\alpha} c_{s,surf}{}^{1-\alpha} c_l^{\alpha} \quad \text{(S22)}$$

where $c_{s,\max}$ and $c_{s,surf}$ denote the maximum and surface Li concentration of active particles, respectively, and $c_l$ is the Li salt concentration of the electrolyte, $k$ the reaction constant. Substituting the stoichiometry $x = c_{s,surf}/c_{s,\max}$ of Li in $Li_xNi_{0.5}Mn_{0.3}Co_{0.2}O_2$ into Equation S22 (there was no concentration gradient in the active particles at the moment a current is loaded) and rewriting Equation S50 with $\alpha$ = 0.5, then we have

$$\eta_{ct} = \frac{2RT}{F} \operatorname{arcsin h} \left( \frac{i}{2Fkc_{s,\max} (1-x)^{0.5} x^{0.5} c_l^{0.5}} \right) \quad \text{(S23)}$$

However, the Butler-Volmer model reveals deviation from the concentration-dependent relation (Figure S10b), though obtaining the same reaction constant ($k = 3\times10^{-11}$ $m^{2.5}$ $mol^{-0.5}$ $s^{-1}$) by fitting the data in Figure S10a,b. For accurate overpotential calculation in the electrochemical simulations, an interpolation function model (Equation S24-S28) was established to approximately depict the relationship among $i$, $\eta_{ct}$, and $x$ (Figure S10c). Additionally, the current-overpotential relation of Li metal electrodes also obeys the Butler-Volmer model in the overpotential range of 0-140 mV (Figure S10d).

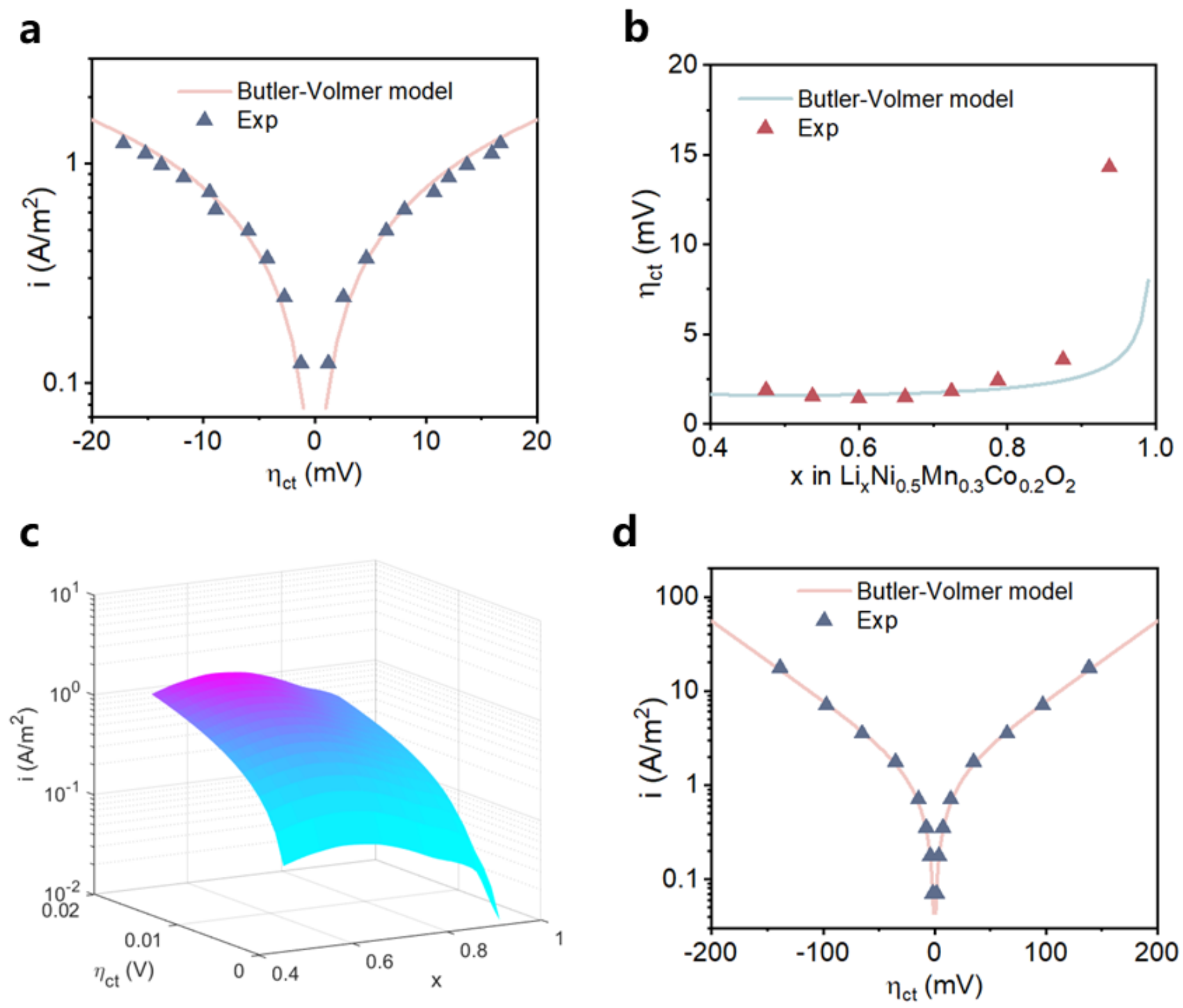

**Figure S10** a,b, Measured overpotentials of charge transfer at various current densities (a) and Li stoichiometry of NMC532 (b) in the three-electrode configuration. c, The plot of the interpolation function model of reaction kinetics created by (a), (b) and Equation S24-S28. d, Current density-overpotential relation of Li metal electrodes.

## Interpolation function model of reaction kinetics

A universal form of electrochemical reaction kinetics can be written as

$$i = f\left(x, \eta_{ct}\right) \tag{S24}$$

Given a fixed current density $i_a$, we can measure the explicit relationship between $x$ and $\eta_{ct}$

$$\eta_{ct} = g\left(x\right), \quad i = i_a \tag{S25}$$

For a specific point, $x = x_0$, the function can be normalized as

$$\frac{\eta_{ct}}{\eta_{ct}\big|_{x=x_0}} = \overline{\eta_{ct}} = \overline{g\left(x\right)} \tag{S26}$$

We assume that the relation Equation S26 is independent of current density $i$, as an approximate treatment. Meanwhile, the current density-overpotential relation at $x = x_0$ can be obtained by experimental measurements,

$$i = f\left(x_0, \eta_{ct}\right) = h\left(\eta_{ct}\big|_{x=x_0}\right), \quad x = x_0 \tag{S27}$$

By substituting Equation S26 into S27, then we have

$$i = h\left(\eta_{ct} \big/ \overline{g\left(x\right)}\right) \tag{S28}$$

## Diffusion coefficients used in the electrochemical simulations

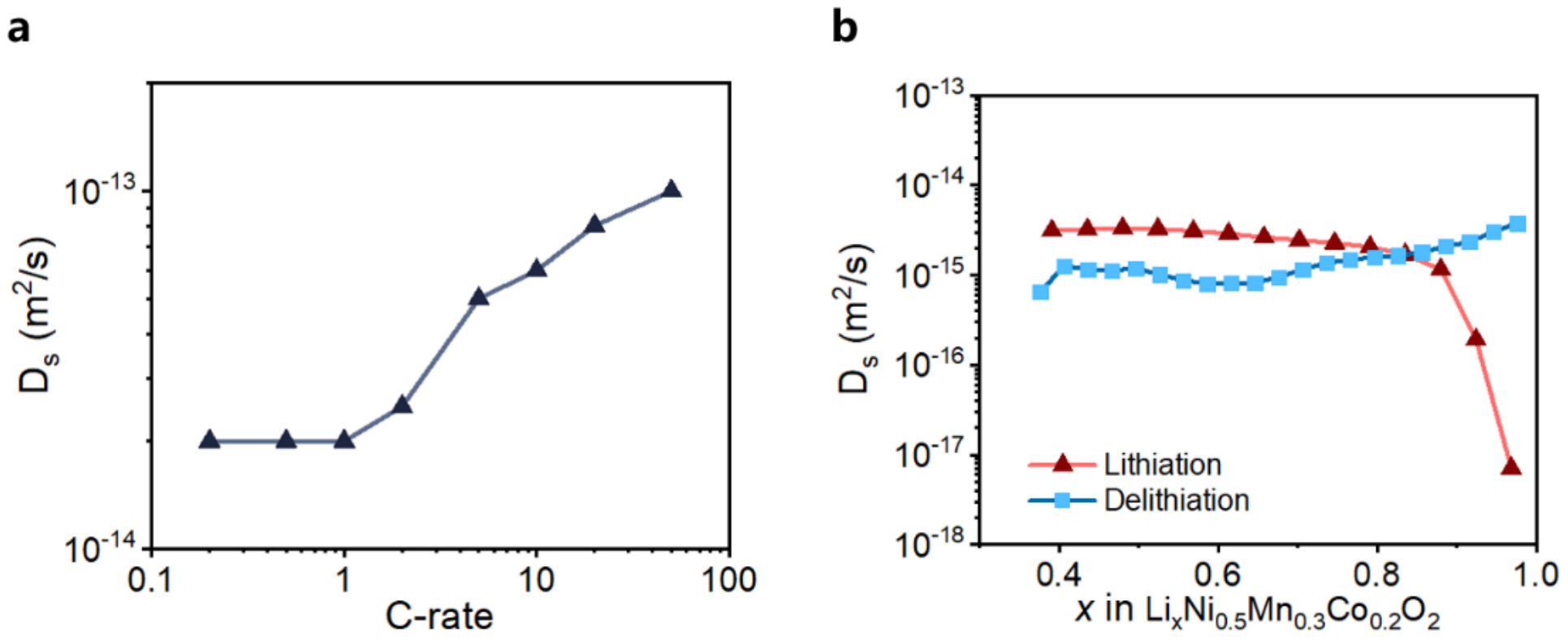

**Figure S11** Apparent diffusion coefficients estimated by fitting the overpotential data with Equations S9 and S53 (a) and measured by GITT testing (b).

The apparent diffusion coefficients (Figure S10a) $D_{s,0}$(C) were estimated by fitting the overpotential data using the Fickian diffusion model (Equation S4) with a concentration-dependent property from GITT (Figure S10b). The diffusion coefficient $D_s$ used in the Fickian diffusion model (Equation S4) is the function of $D_{s,app}$ and the measured diffusivity $D_{s,GITT}$ by GITT, as expressed by:

$$D_s\left(x,C\right)=D_{s,0}(C)\frac{D_{s,GITT}\left(x\right)}{\max\limits_{x}\left\{D_{s,GITT}\left(x\right)\right\}} \tag{S29}$$

The rate dependence of $D_{s,0}$(C) may be attributed to non-Fickian diffusion and also found in Ref. [7,8].

## Comparison of the simulation results from different SLPE models

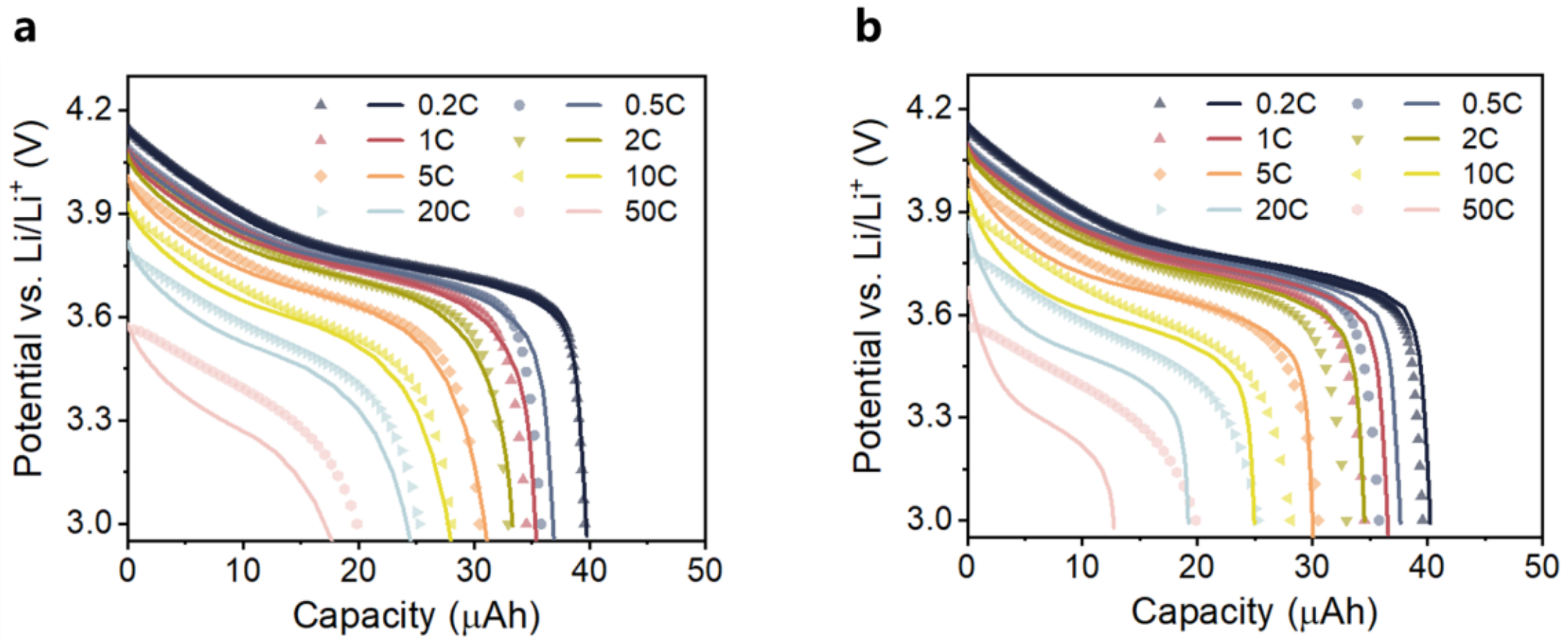

**Figure S12** Electrochemical simulation results compared with the experimental voltage data of SLPE cells. (a) The interpolation function model with a concentration- and rate-dependent diffusivity and (b) the Bulter-Volmer model with a concentration-independent diffusivity, respectively.

## Electrochemical model of CE cells and overpotential calculations

**Model assumptions.** The electrochemical model of CE cells is developed based on Newman's porous electrode theory [1,2], while the reaction kinetics of Li intercalation is controlled by the interpolation function model (Equation S9) in this work. The CE cell model simplifies the CE as a continuum medium by the volume average method, and only considers the kinetic processes in the electrode thickness direction by assuming a uniform electrode plane. The porous electrode and the electrolyte are superimposed in space and interact via the interface reaction.

**Governing equations.** In the CE domain, the NMC532 particles are assumed to be ideally spherical and isometrical, and the Li diffusion process is also described by Equation S4 in a spherical coordinate:

$$\frac{\partial c_s}{\partial t} = D_s\left(\frac{\partial^2 c_s}{\partial r^2} + \frac{2}{r}\cdot\frac{\partial c_s}{\partial r}\right) \tag{S30}$$

Similar to Equation S5, the potential distribution on the porous electrode is governed by:

$$-\sigma_{eff}\frac{\partial^2 \phi_s}{\partial x^2} = a_s F i \tag{S31}$$

where $a_s$ is the specific surface of active particles, $\sigma_{eff}$ is the effect electrode conductivity, calculated by

$$\sigma_{eff} = \frac{\varepsilon_{ap}}{\tau_{ap}}\sigma \tag{S32}$$

where $\varepsilon_{ap}$ and $\tau_{ap}$ are the porosity and tortuosity of the active particles, respectively.

The transport process of $Li^+$ ions in the electrolyte domain is governed by:

$$\varepsilon\frac{\partial c_l}{\partial t} = \frac{\partial}{\partial x}\cdot\left(D_{l,eff}\frac{\partial c_l}{\partial x}\right) + \frac{(1-t_+)a_s i}{v_+} \tag{S33}$$

where $\varepsilon$ is the electrode porosity. For the charge transfer process in the electrolyte, it is governed by:

$$-\frac{\partial}{\partial x}\cdot\left(\kappa_{eff}\frac{\partial\phi_l}{\partial x}\right)+\frac{2RT(1-t_+)}{F}\left(1+\frac{\partial\ln f_\pm}{\partial\ln c_l}\right)\frac{\partial}{\partial x}\cdot\left(\kappa_{eff}\frac{\partial\ln c_l}{\partial x}\right)=a_sFi \qquad \text{(S34)}$$

where $D_l$ and $\kappa$ are modified by $\varepsilon$ and the pore tortuosity $\tau$as Equation S8 in the electrolyte domain.

The electrochemical reactions, contact resistance, and film resistance of the CEs are also described by Equation S9-S13.

**Boundary conditions.** The boundary conditions of Li diffusion within the active particles are:

$$-D_s\frac{\partial c_s}{\partial r}\bigg|_{r=0}=0,\ -D_s\frac{\partial c_s}{\partial r}\bigg|_{r=R_{ap}}=i \qquad \text{(S35)}$$

For the electron conduction in the porous electrode, the boundary conditions are

$$\frac{\partial\phi_s}{\partial x}=0 \qquad \text{(S36)}$$

at the electrode-separator interface and

$$-\sigma_{eff}\frac{\partial\phi_s}{\partial x}=\frac{I}{A} \qquad \text{(S37)}$$

at the cathode-current collector interface, where $I$ is the total current applied to the CE cell, $A$ the electrode area.

The boundary conditions of the species transport in the electrolyte solution are

$$\frac{\partial c_l}{\partial x}=0,\quad -D_{l,eff}\frac{\partial c_l}{\partial x}=\frac{i_{Li}}{F} \qquad \text{(S38)}$$

for the current collector surface and the Li metal surface, respectively. The boundary conditions of the charge transport are

$$\frac{\partial\phi_l}{\partial x}=0,\quad -\kappa_{eff}\frac{\partial\phi_l}{\partial x}=i_{Li} \qquad \text{(S39)}$$

for the current collector surface and the Li metal surface, respectively.

The electrostatic potential of the Li counter electrode is also set as Equation S20.

**Overpotential calculations.** Different from the SLPE system, the overpotential calculation is much more complicated for the CE cells due to the gradient distributions of the potentials ($\phi_s$, $\phi_l$) and species concentrations ($c_s$, $c_l$) in the thickness direction, which means that the overpotential from each kinetic process is position-dependent. To sufficiently reflect the electrolyte polarization, the point at the electrode-current collector interface is selected since the electron resistance through the composite electrode is relatively small (Figure S12). For simplicity, all the overpotentials are positive for discharging.

The overpotential produced in the separator domain is the sum of the Ohm part and the concentration polarization part

$$\eta_{sep} = \eta_{sep,\Omega} + \eta_{sep,c} \tag{S40}$$

where

$$\eta_{sep,\Omega} = \int_0^{L_{sep}} \frac{i_l}{\kappa_{eff,0}} dx \tag{S41}$$

and

$$\eta_{sep,c} = \int_0^{L_{sep}} i_l \left( \frac{1}{\kappa_{eff}} - \frac{1}{\kappa_{eff,0}} \right) - \frac{2RT(1-t_+)}{F} \left( 1 + \frac{\partial \ln f_\pm}{\partial \ln c_l} \right) \frac{\partial \ln c_l}{\partial x} dx \tag{S42}$$

Likewise, the overpotential produced in the electrode pores can be expressed as

$$\eta_l = \eta_{l,\Omega} + \eta_{l,c} \tag{S43}$$

where

$$\eta_{l,\Omega} = \int_{L_{sep}}^{L_{sep}+L_{pos}} \frac{i_l}{\kappa_{eff,0}} dx \tag{S44}$$

and

$$\begin{aligned} \eta_{l,c} = \int_{L_{sep}}^{L_{sep}+L_{pos}} & i_l \left( \frac{1}{\kappa_{eff}} - \frac{1}{\kappa_{eff,0}} \right) \\ & - \frac{2RT(1-t_+)}{F} \left( 1 + \frac{\partial \ln f_\pm}{\partial \ln c_l} \right) \frac{\partial \ln c_l}{\partial x} dx \end{aligned} \tag{S45}$$

The concentration overpotential caused by the Li concentration gradient within the active particles can be calculated by

$$\eta_s = E_{eq,ave} - E_{eq,surf} \tag{S46}$$

The reaction overpotential of charge transfer at the particle-electrolyte interface can be obtained according to Equation S9

$$\eta_{ct} = \overline{g\left(c_{s,surf}/c_{s,\max}\right)} h^{-1}\left(i/\left(c_l/c_{l,0}\right)^{0.5}\right) \tag{S47}$$

At the Li metal-electrolyte interface, it is

$$\eta_{Li,ct} = \frac{2RT}{F} \operatorname{arcsin h}\left(\frac{i_{Li}}{2Fkc_l^{0.5}}\right) \tag{S48}$$

The overpotential from SEI film is

$$\eta_{Li,sei} = IR_{sei} \tag{S49}$$

Therefore, the total overpotential from the Li metal anode is

$$\eta_{Li} = \eta_{Li,ct} + \eta_{Li,sei} \tag{S50}$$

The electron conduction resistance on the electrode matrix, adding the contact resistance at the electrode-current collector interface, contributes to the Ohm's voltage drop

$$\eta_e = IR_e \tag{S51}$$

The output voltage of a CE cell hence can be expressed as

$$V_{cell} = E_{eq,ave} - \left(\eta_l + \eta_s + \eta_{ct} + \eta_e\right) - \eta_{sep} - \eta_{Li} \tag{S52}$$

## Electrochemical inhomogeneity across the electrode depth

It should be noted that the overpotentials are non-uniform in the direction of electrode thickness, due to the gradient distributions of electrostatic potentials and species concentrations (Figure S13). It means different overpotential components in each route of such a parallel circuit structure [9]. Specifically, three representative lines are illustrated in Figure S13, which highlight $\eta_e$ (the $e^-$-dominated line) or $\eta_l$ (the $Li^+$-dominated line), or balances $\eta_e$ and $\eta_l$ (the $e^-$-$Li^+$ balanced line). Herein, the $Li^+$-dominated line was selected to sufficiently reflect the electrolyte polarization in the situation that the electron resistances through the CEs are relatively small (Figure S14b and Figure S17).

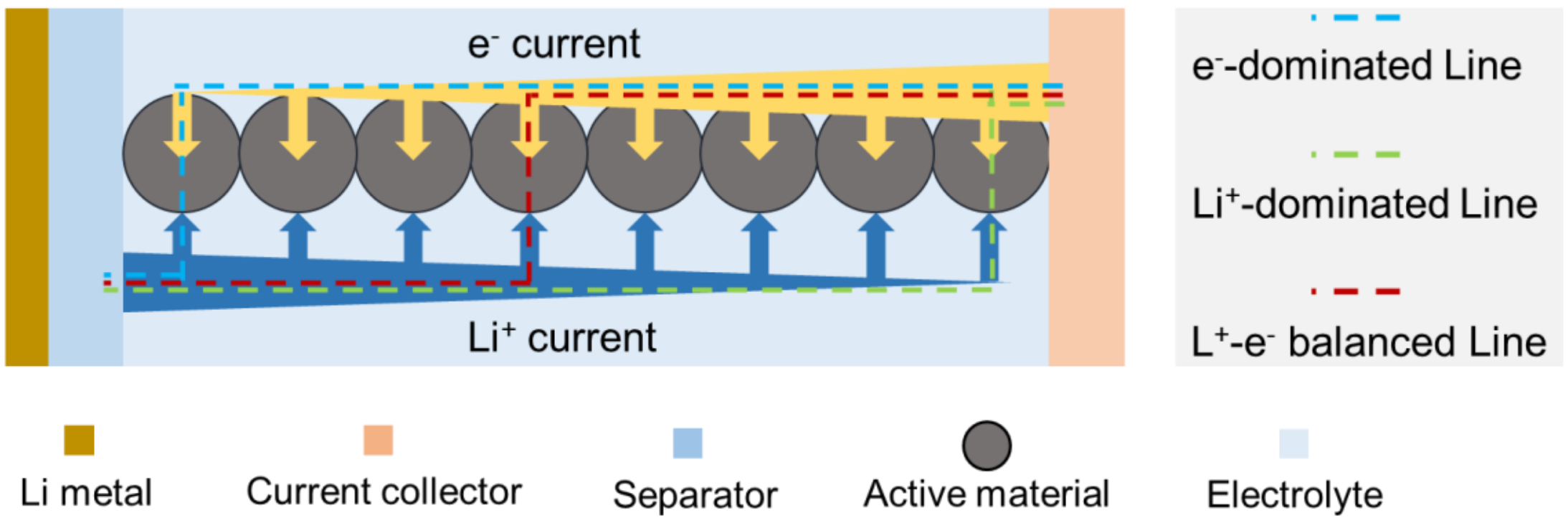

**Figure S13** Schema of $e^-$ lines (yellow arrows) and $Li^+$ lines (blue arrows) in the circuit of CEs and the calculation routes of overpotentials (dashed lines).

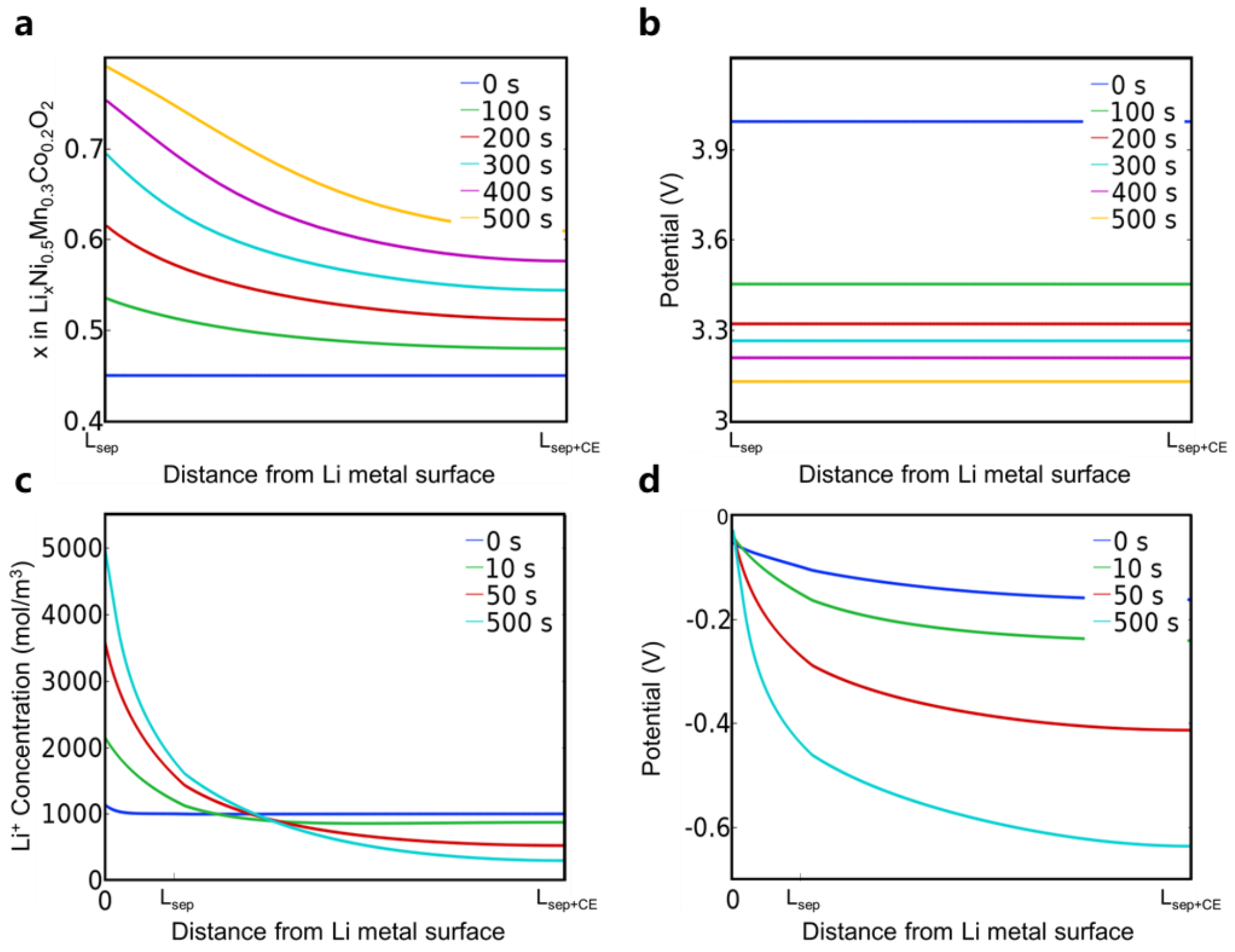

**Figure S14** Gradient distributions of potentials and species concentrations in the CE cell with Design 1 during discharging at 3C rate. a,b, Li stoichiometry (a) and potential (b) of the solid electrode. c,d, $Li^+$ concentration (c) and potential (d) of the electrolyte.

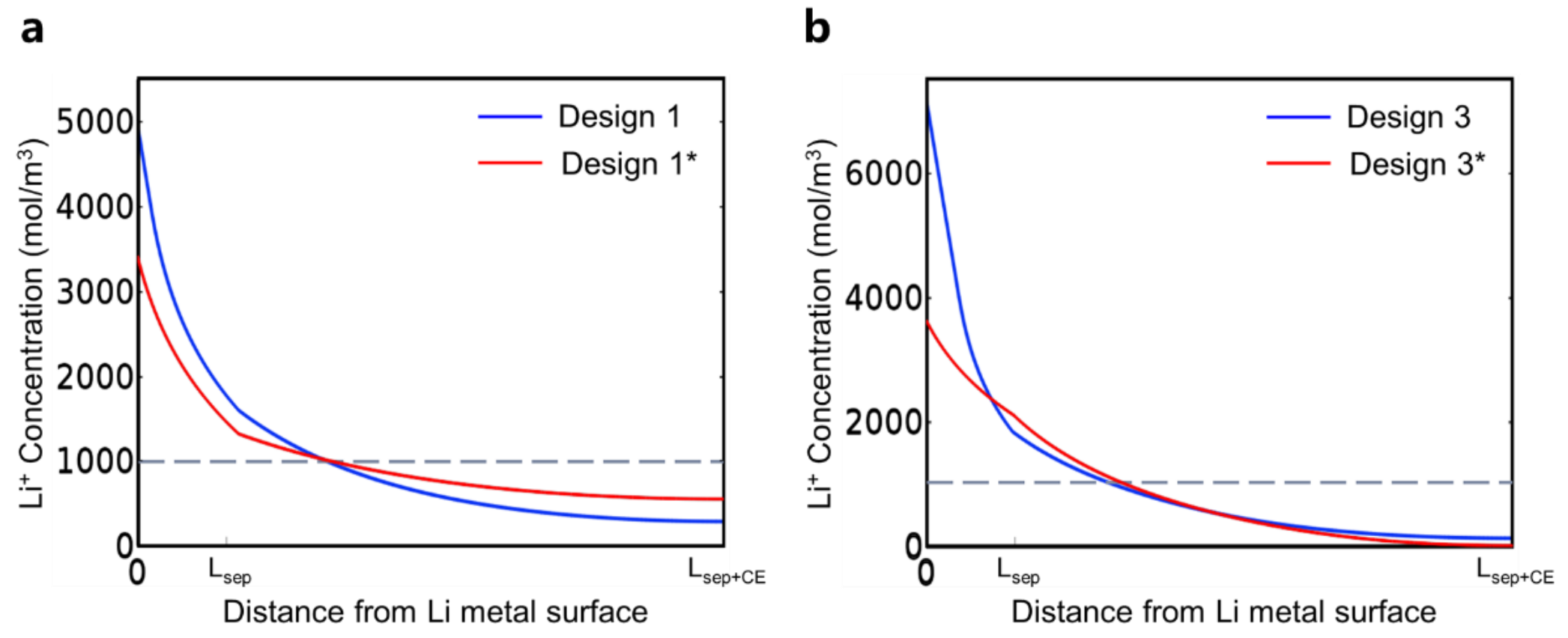

**Figure S15** Gradient distributions of $Li^+$ concentrations in the CE cells at 500s of 3C discharging. a, Design 1 (the original with $\tau_{eff}$=2.8) and Design 1* (the optimized with $\tau_{eff}$=1.6). b, Design 3 (the original with $N_M$=10.4) and Design 3* (the optimized with $N_M$=4.4).

# Model validation

### Table S2

Manufacture information of the experimental CE cells in three designs of CEs.

| Parameters | Design 1 | Design 2 | Design 3 |
|---|---|---|---|
| Areal capacity (mAh/cm$^2$) | 4 | 3 | 4.6 |
| Electrode thickness (μm) | 120 | 119 | 145 |
| Fraction of active material | 0.49 | 0.36 | 0.46 |
| Electrode porosity | 0.39 | 0.55 | 0.42 |

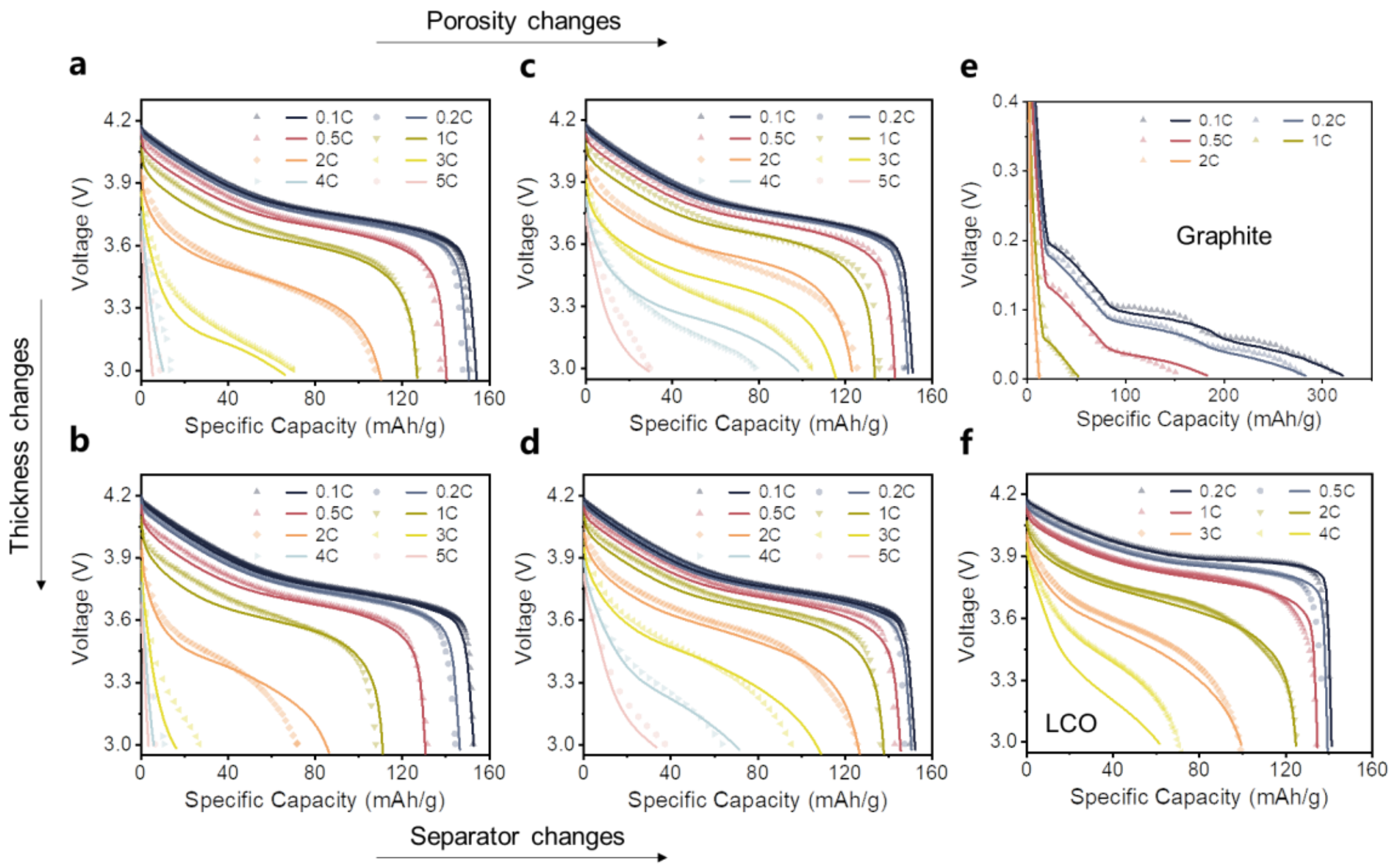

**Figure S16** Simulation results (solid lines) compared with experimentally measured voltage (dots) of CE cells at varied C-rates. NMC532: Design 1 (a), Design 3 with $N_M$=10.4 (b), Design 2 (c), and Design 3 with $N_M$=4.4 (d); graphite (e); LCO (f).

# EIS of the CEs in different designs

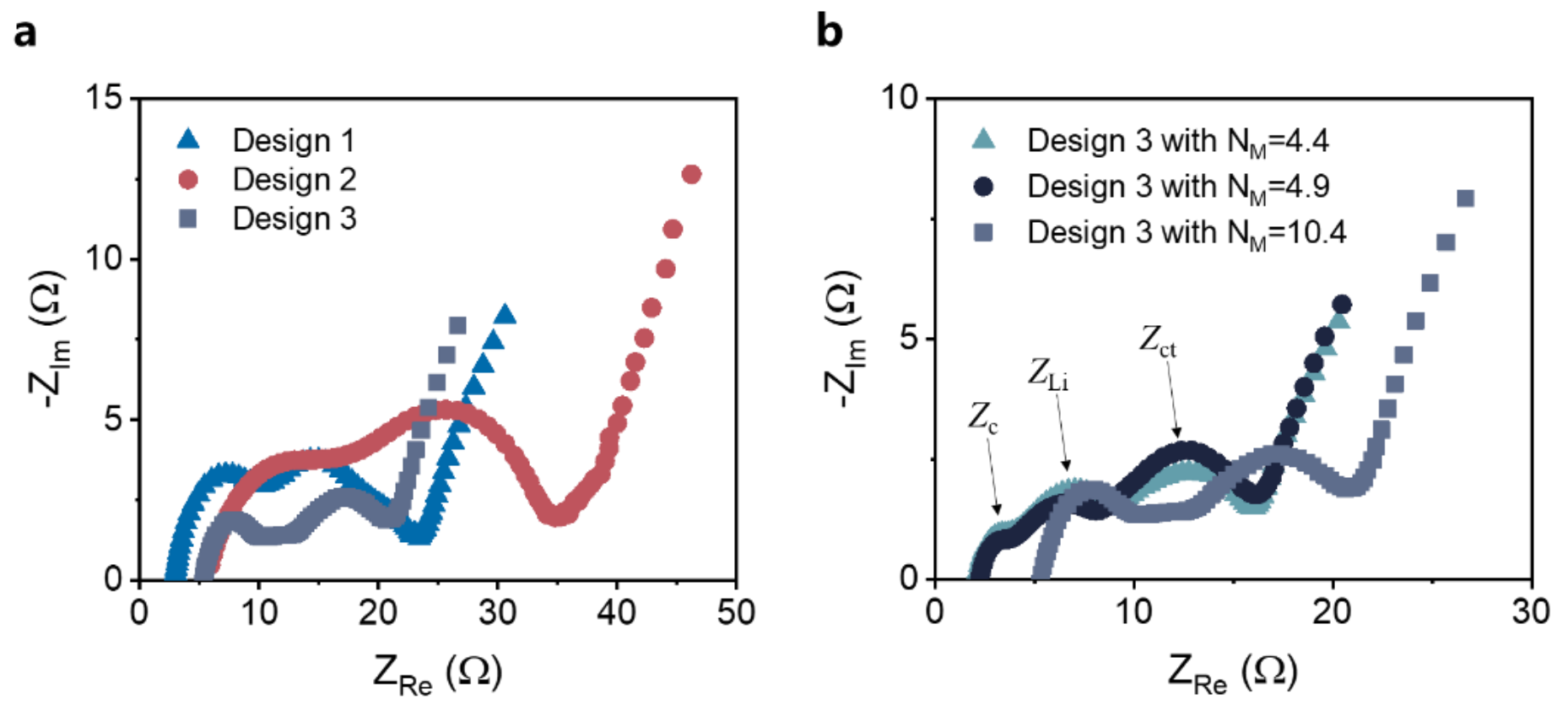

**Figure S17** EIS of the CEs with different electrode designs (a) and separators (b).

## Separator parameters

The separator tortuosity was calculated by Equation S53 which is easily derivated from Equation S8 and Ohm's law.

$$\tau_{sep} = \frac{\varepsilon_{sep} \kappa R_{\Omega} A}{L_{sep}} \quad \text{(S53)}$$

$R_{\Omega}$ were obtained from the intercepts of the EIS curves with the real axis in the Nyquist plot, as shown in Figure S18. The porosity, tortuosity, and MacMullin number of the Celgard separators are listed in Table S3, and the values are in agreement with the reported results in Ref. [10].

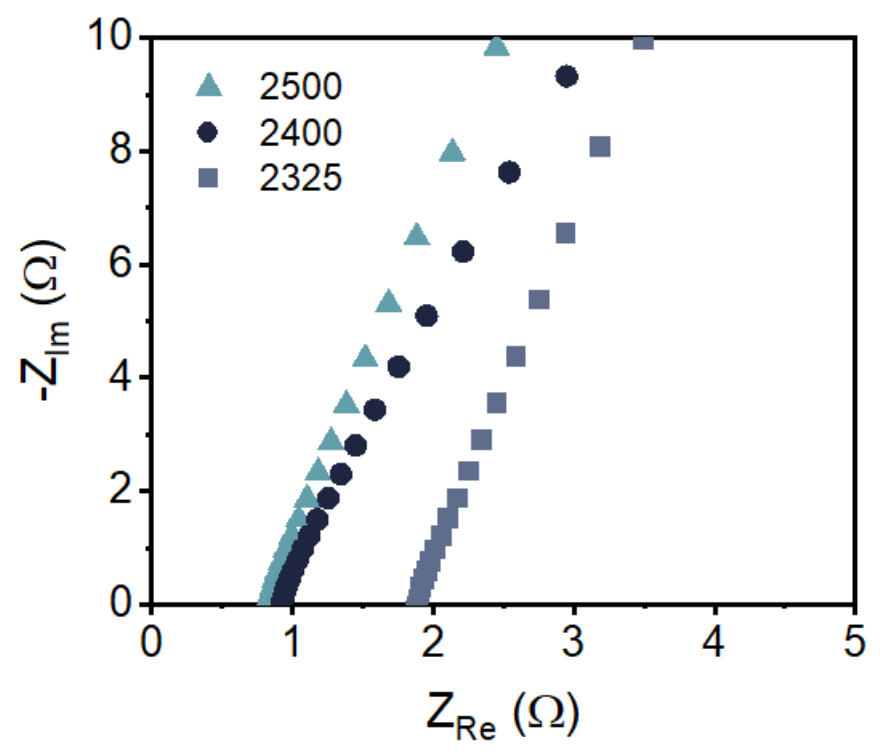

**Figure S18** EIS of the commercial separators (Celgard 2325, 2400, 2500) soaked in the electrolyte solution.

**Table S3**

Separator parameters from manufacturer and EIS measurements.

| Separator parameters | Celgard 2325 | Celgard 2400 | Celgard 2500 |
|---|---|---|---|
| Thickness (μm) | 25 | 25 | 25 |
| Porosity | 0.39 | 0.41 | 0.55 |
| Tortuosity | 4.1 | 2 | 2.4 |
| MacMullin number | 10.4 | 4.9 | 4.4 |